\documentclass{aa}  

\bibpunct{(}{)}{;}{a}{}{,} 

\usepackage[english]{babel}
\usepackage[utf8]{inputenc}
\usepackage[T1]{fontenc}
\usepackage[]{tocloft}
\usepackage{lipsum}

\usepackage{amsmath}
\usepackage{amssymb}
\usepackage{amsfonts}
\usepackage{mathtools}
\usepackage{booktabs}
\usepackage{longtable}
\usepackage{stmaryrd}
\usepackage{mathrsfs}
\usepackage{multirow}
\usepackage{centernot}
\usepackage{placeins}

\usepackage{fancyvrb}
\usepackage{graphicx}
\usepackage{subfig}
\usepackage{wrapfig}
\usepackage[usenames,dvipsnames,table]{xcolor}
\usepackage{cases}
\usepackage{multicol}
\usepackage{tikz}
\usepackage{eurosym} \def\{\euro{}} 
\usepackage{cmlgc}
\usepackage{color}
\usepackage{empheq}
\usepackage[most]{tcolorbox}
\usepackage{booktabs}
\usepackage{textcomp} 
\usepackage{soul}

\usepackage{graphicx}
\usepackage{txfonts}
\usepackage[toc]{appendix}
\usepackage{url}
\definecolor{linkcolor}{rgb}{0,0,0}
\definecolor{linkcolorurl}{rgb}{0,0,1}
\usepackage[colorlinks=true,
linkcolor= BrickRed,
citecolor= BrickRed,
urlcolor= BrickRed]
{hyperref}

\newcommand{\para}[1]{{\noindent \textbf{#1} \,}}

\input{Commandes.sty}

\defcitealias{Crinquand_2020}{C20}

\begin{document}

   \title{Synthetic gamma-ray light curves of Kerr black hole magnetospheric activity from particle-in-cell simulations}
   
   \author{B. Crinquand \inst{1}
          \and
          B. Cerutti \inst{1}
          \and
          G. Dubus \inst{1}
          \and
          K. Parfrey \inst{2}
          \and
          A. Philippov \inst{3}
          }
    

   \institute{Univ. Grenoble Alpes, CNRS, IPAG, 38000 Grenoble, France \\
   \email{benjamin.crinquand@univ-grenoble-alpes.fr}
              \and 
              School of Mathematics, Trinity College Dublin, Dublin 2, Ireland
              \and
              Center for Computational Astrophysics, Flatiron Institute, 162 Fifth Avenue, New York, NY 10010, USA
              }
              

   \date{Received ... /
         Accepted ...}
         

 
  \abstract
  {The origin of ultra-rapid flares of very high-energy radiation from active galactic nuclei remains elusive. Magnetospheric processes, occurring in the close vicinity of the central black hole, could account for these flares.}
  {Our aim is to bridge the gap between simulations and observations by synthesizing gamma-ray light curves in order to characterize the activity of a black hole magnetosphere, using kinetic simulations.}
  {We performed global axisymmetric 2D general-relativistic particle-in-cell simulations of a Kerr black hole magnetosphere. We included a self-consistent treatment of radiative processes and plasma supply, as well as a realistic magnetic configuration, with a large-scale equatorial current sheet. We coupled our particle-in-cell code with a ray-tracing algorithm in order to produce synthetic light curves.}
  {These simulations show a highly dynamic magnetosphere, as well as very efficient dissipation of the magnetic energy. An external supply of magnetic flux is found to maintain the magnetosphere in a dynamic state, otherwise the magnetosphere settles in a quasi-steady Wald-like configuration. The dissipated energy is mostly converted to gamma-ray photons. The light curves at low viewing angle (face-on) mainly trace the spark gap activity and exhibit high variability. On the other hand, no significant variability is found at high viewing angle (edge-on), where the main contribution comes from the reconnecting current sheet.}
  {We observe that black hole magnetospheres with a current sheet are characterized by a very high radiative efficiency. The typical amplitude of the flares in our simulations is lower than  is detected in active galactic nuclei. These flares could result from the variation in parameters external to the black hole.}


\titlerunning{Synthetic gamma-ray light curves of Kerr black hole magnetospheric activity}

   \maketitle
%

\section{Introduction}

Ground-based Cherenkov telescopes have shown that active galactic nuclei (AGN) can be highly variable sources of very high-energy (VHE) emission ($>100$ GeV) \citep{Aharonian_2007,Albert_2007,Aleksic_2014}. Variability timescales can be shorter than the horizon light-crossing time $t_\mathrm{g} = r_\mathrm{g} / c$, where $r_\mathrm{g}$ is the gravitational radius of the central supermassive black hole. This constrains emission models, as the size of the emitting region must be  on the order of $r_\mathrm{g}$ by virtue of causality. 



This variability was primarily observed in blazars, AGN with jets lying close to our line of sight. However, TeV $\gamma$-ray flares were also detected from the nuclei of the radio galaxies M87 \citep{Aharonian_2006,Aliu_2012} and Centaurus A~\citep{Aharonian_2009} for instance, these galaxies having jets misaligned with our line of sight by more than $\simeq 15^\circ$. This suggests that VHE flares are a widespread feature in AGN. The variability timescale $\Delta t = 2$ days of M87* VHE flares is comparable to the horizon light-crossing time $t_\mathrm{g} \simeq 0.4$ days. M87 is of paramount importance, and has attracted considerable attention because it is close enough that its nucleus can be resolved by radio interferometry~\citep{EHT_1}. This has allowed observers to establish a connection between VHE flares and a brightening of the radio core~\citep{Acciari_2009}, which occurred simultaneously on several occasions. These observations also ruled out other potential compact VHE emission sites, such as knots in the relativistic jet. Thus, we might be able to link the formation of such a jet with processes at play in the close vicinity of the central black hole.

The extreme variability of these flares challenges conventional models of AGN~\citep{Rieger_2012}. Because this emission seems to originate from the vicinity of the black hole, we are motivated to study  nonthermal magnetospheric processes as a possible source~\citep{Katsoulakos_2018,Levinson_2011,Hirotani2016,Levinson_2000}. This model is applicable to low-luminosity AGN, such as M87* or Sgr A*. If the luminosity of the accretion flow is low enough, the plasma density can drop below the Goldreich-Julian value, which is required to screen the electric fields generated by the dragging of the magnetic field lines by the black hole~\citep{Wald_1974}. Consequently, spark gaps arise, accelerating particles to very high energies; these energetic particles then scatter off soft photons from the accretion flow to produce VHE emission. Subsequent pair production is triggered by the annihilation of TeV photons with soft photons. This fresh plasma screens the electric field, quenching particle acceleration and nonthermal radiation. As the plasma inevitably flows away from the gap, the electric field is restored and VHE emission resumes. Hence, depending on the gap size, this model may account for the variability of the VHE emission observed from AGN. An electromagnetic cascade takes hold, feeding on the black hole to produce VHE emission and pair plasma. An important takeaway from this model is the possibility to activate the Blandford-Znajek (BZ) process by providing the plasma necessary to establish a quasi-force-free magnetosphere~\citep{Blandford_1977}. Since the presence of a gap has an impact on the global structure of the magnetosphere, and because the electromagnetic cascade is a highly nonlinear phenomenon, numerical simulations are well suited to gain insight into this problem.

\citet{Parfrey_2019} demonstrated the feasibility of performing global general relativistic particle-in-cell (GRPIC) simulations of a black hole magnetosphere. They modeled a black hole immersed in an initially vertical magnetic field, but used a simplified treatment for plasma supply that could only mimic how an electromagnetic cascade develops. This was improved upon in \citealt{Levinson_2018}, \citealt{Chen_2020}, and \citealt{Crinquand_2020}, where a self-consistent treatment of inverse Compton (IC) scattering and pair production was implemented. In \citet[][hereafter C20]{Crinquand_2020}, we simulated a monopole magnetosphere to capture the intrinsic activity of spark gaps, and showed that the BZ process could be successfully activated, as the magnetosphere was filled with pair plasma produced in the ergosphere. The size of the gap was consistent with sub-horizon variability. 

However, in the case of isolated magnetospheres, more realistic configurations with a large-scale poloidal magnetic field should display an equatorial current sheet~\citep{Komissarov_2004a,Komissarov_2007}. This current sheet would originate from the need to close the electric current system, since negative currents flow from both poles (if the spin axis is aligned with the magnetic field). Such a situation can come up if the accretion flow is truncated at large radius, causing the  accretion to pause for a while, as can happen in magnetically arrested disk simulations~\citep{Narayan_2003}. Still, large-scale and intermittent ergospheric current sheets are expected to develop naturally in accreting black hole magnetospheres as well \citep[e.g.,][]{Ripperda_2020}, highlighting the need to understand their importance. Magnetic reconnection, an intermittent phenomenon that is known to accelerate particles very efficiently, is ubiquitous in such current sheets~\citep{Kagan_2015}. It could also be responsible for variable VHE emission~\citep{Cerutti_2012,Christie_2019,Mehlhaff_2020}. It is unclear how such a current sheet can affect the pair discharge mechanism, and what the relative contributions of the polar cap and the current sheet emissions are. In this paper we extend the work carried out in \citetalias{Crinquand_2020}, no longer neglecting the equatorial reconnection activity. Now that the polar cap activity has been characterized, we can aim toward  more realistic magnetic configurations. In addition, contrary to previous kinetic studies, here we make a point of extracting physical observables from our simulations by implementing a more efficient treatment of ray tracing. This allows us to synthesize gamma-ray light curves, assimilating  HE and VHE emission to IC processes.

This paper is divided into two main sections. In Sect.~\ref{sec1} we describe our numerical scheme and setup, and present our new simulations of magnetospheric activity. In Sect.~\ref{sec2} we focus on producing observables from PIC simulations. We present synthetic light curves for the simulations carried out in Sect.~\ref{sec1} and in \citetalias{Crinquand_2020}. The post-process treatment is described in Appendix~\ref{appendix}.





\section{Magnetospheric activity} \label{sec1}

\subsection{Numerical techniques} \label{sec:techniques}

In this section only we set $c=1$. \\

\para{Metric} In this work we perform 2D global GRPIC simulations, using a general relativistic version of the PIC code \texttt{Zeltron}~\citep{Cerutti_2013,Cerutti_2015}, first introduced in \citet{Parfrey_2019}. The background spacetime is described by the Kerr metric, with a spin parameter $a \in [0,1[$. The code uses spherical Kerr-Schild coordinates $(t,r,\theta,\varphi)$, which are not singular at the event horizon (see~\citealt{Komissarov_2004a} for an expression of the coefficients of the metric). We use the 3+1 formulation of general relativity \citep{MacDonald_1982} in order to evolve the particles and fields with respect to a universal coordinate time $t$. This formulation naturally introduces fiducial observers (FIDOs), whose wordlines are orthogonal to the spatial hypersurfaces of constant $t$ as privileged observers. In an axisymmetric and stationary spacetime such as the one described by the Kerr metric, they are also zero angular momentum observers (ZAMOs). In this formulation, the metric can be rewritten such that the line element $\diff{s^2}$ reads

\begin{equation}
\diff{s^2} = - \alpha^2 \mathrm{d}t^2 + h_{ij} (\mathrm{d}x^i + \beta^i \mathrm{d}t) (\mathrm{d}x^j + \beta^j \mathrm{d}t).
\end{equation}
In this equation $\alpha$ is the lapse function (the redshift of a FIDO with respect to the coordinate time $t$), $\vec{\beta}$ is the shift vector (the $3$-velocity of a FIDO with respect to the coordinate grid), whereas $h_{ij}$ denotes the spatial $3$-metric associated with the spatial hypersurfaces of constant $t$. The gravitational radius of the black hole is denoted $r_\mathrm{g}$.
\\


\para{Electromagnetic fields} We solve the electromagnetic field equations derived by \citet{Komissarov_2004a}
\begin{eqnarray}
    \partial_t \vec{B} & = & - \Nabla \times \vec{E} \label{eq:maxwell_B},\\
    \partial_t \vec{D} & = & \Nabla \times \vec{H} - 4 \pi \vec{J}, \label{eq:maxwell_D}
\end{eqnarray}
where $\vec{B}$ and $\vec{D}$ are respectively the magnetic and electric fields locally measured by FIDOs (they are physical observables), and  $\vec{H}$ and $\vec{E}$ are auxiliary fields defined by
\begin{eqnarray}
    \vec{H} & = & \alpha \vec{B} - \vec{\beta} \times \vec{D}, \\
    \vec{E} & = & \alpha \vec{D} + \vec{\beta} \times \vec{B}.
\end{eqnarray}
The auxiliary current density $\vec{J}$ is related to the electric current density measured by FIDOs $\vec{j}$ via $\vec{J}=\alpha \vec{j} - \rho \vec{\beta}$, $\rho$ being the electric charge density measured by FIDOs. These fields are defined on a spatial Yee grid~\citep{Yee_1966}. Equations~\eqref{eq:maxwell_B} and~\eqref{eq:maxwell_D} resemble classical Maxwell-Amp\`ere and Maxwell-Faraday equations, so they can be solved by classic finite-difference time-domain schemes, with additional steps due to the introduction of the auxiliary fields. Since the Maxwell-Gauss equation $\Nabla \cdot \vec{D} = 4 \pi \rho$ is not enforced by the code, we have to regularly perform divergence cleaning~\citep{Cerutti_2015}.
\\

\para{Particles} We simulate pair plasma. In the $3+1$ formalism, the Hamiltonian of a positron (or electron) of charge $q=\pm e$, mass $m_\mathrm{e}$ and $4$-velocity $u_\mu$, moving in an electromagnetic field with $4$-potential $A_\mu$, reads
\begin{equation} \label{eq:hamiltonian}
    \mathcal{H} (x^j,u_i) = - u_0 = \alpha \sqrt{1+ h^{jk} u_j u_k} - \beta^i u_i - \dfrac{q}{m_\mathrm{e}} A_0.
\end{equation}
The particle's equations of motion are deduced from Eq.~\eqref{eq:hamiltonian} and Hamilton's equations~\citep{Hughes_1994,Dodin_2010,Bacchini_2018_1,Bacchini_2018_2, Parfrey_2019}:
\begin{align}
        \ddroit{x^i}{t} = & \, \drond{\mathcal{H}}{u_i} = v^i = \dfrac{\alpha}{\Gamma} h^{ij} u_j - \beta^i, \label{eq:motion_x}\\
    \ddroit{u_i}{t} = & - \drond{\mathcal{H}}{x^i} = - \Gamma \partial_i \alpha + u_j \partial_i \beta^j - \dfrac{\alpha}{2 \Gamma} u_j u_k \partial_i h^{jk} + \dfrac{\alpha}{m_\mathrm{e}} F_i, \label{eq:motion_u}
\end{align}
where $\vec{F}=q(\vec{D}+ (\vec{u}/\Gamma) \times \vec{B})$ is the Lorentz force, $\Gamma = \sqrt{1+ h^{jk} u_j u_k}$ is the FIDO-measured Lorentz factor of the particle, $\vec{v}$ is its $3$-velocity with respect to the grid, and $\vec{u}/\Gamma$ its FIDO-measured $3$-velocity. The electric current density $\vec{J}$ source term involved in Eq.~\eqref{eq:maxwell_D} is determined by the contributions $q \vec{v}$ of each particle.
\\

\para{Photons} In order to treat plasma supply self-consistently in our simulations, we use the radiative transfer algorithm introduced in~\citet{Levinson_2018} and \citetalias{Crinquand_2020} (see the Supplemental Material therein for more details) in order to include IC scattering and $\gamma\gamma$ pair production. We include high-energy photons as a neutral third species that propagates along null geodesics. All particles evolve in a background soft radiation field, putatively emitted by the accretion flow, which makes the propagating medium opaque. We assume that this radiation field is static, homogeneous (with uniform density $n_0$ in any FIDO frame), isotropic, and mono-energetic (with energy $\varepsilon_0$). The opacity of the medium for all particles is parameterized by the fiducial optical depth $\tau_0= n_0 \sigma_\mathrm{T} r_\mathrm{g}$, where $\sigma_\mathrm{T}$ is the Thomson cross-section. At every time step a lepton can scatter off a background soft photon to produce a high-energy photon by IC scattering, whereas a high-energy photon can annihilate with a soft photon to produce an $e^\pm$ pair. 

Two photons of energies $\varepsilon$ and $\varepsilon'$, colliding with an angle $\theta_0$, can only produce a pair provided $\varepsilon \, \varepsilon' \left( 1 - \cos{\theta_0} \right) / 2 \ge (m_\mathrm{e} c^2)^2$ \citep{Gould_1967}. One of our main goals is to characterize high-energy emission by synthesizing light curves from our simulations, that can directly be compared to observations. To do so we need to keep track of the IC photons emitted below the pair-creation threshold $(m_\mathrm{e} c^2)^2 / \varepsilon_0$. These photons are able to escape the soft photon field: they are responsible for the magnetospheric component of the high-energy emission received from Earth. However, these photons do not participate in the plasma dynamics. It is therefore critical to save the information they carry, as  detailed in Sect.~\ref{sec2}, before discarding them from the simulation. We do not include synchrotron and curvature radiation, which will be considered in a future work. \\

\para{Parameters} The simulations are axisymmetric: the particles move in  3D space, with all quantities being invariant by rotation around the spin axis of the black hole. The simulation domain is $r\in [r_\mathrm{min}=0.9\, r_{\rm h}, r_\mathrm{max}=10 \, r_{\rm g}]$, $\theta \in [0,\pi]$, with $r_{\rm h} = r_{\rm g} ( 1+\sqrt{1-a^2} )$ the radius of the event horizon. Fields are defined on a grid evenly spaced in $\log_{10}{(r)}$ and $\theta$. The spin parameter $a$ is fixed at $0.99$. We focus on the high optical depth regime, and run simulations with $\tau_0 = 30$, $50$, and $80$.

The normalized magnetic field is defined by $\tilde{B}_0 = r_{\rm g} (e B_0 / m_\mathrm{e} c^2)$, with $B_0$ the magnetic field strength at the horizon, whereas the normalized radiation field energy is $\tilde{\varepsilon}_0 = \varepsilon_0 / m_\mathrm{e} c^2$. We kept $\tilde{B}_0 = 5\times 10^5$ and $\tilde{\varepsilon}_0=5\times10^{-3}$ fixed in the simulation, in accordance with the criteria described in \citetalias{Crinquand_2020}. This ensures that secondary pair-produced particles have high Lorentz factors, and that primary particles can be accelerated to energies above the pair-creation threshold. The magnetosphere is initially filled with high-energy photons, distributed uniformly and isotropically from $r=r_\mathrm{h}$ up to $r=4 \, r_{\rm g}$, with the energy $\varepsilon_1=200 \, m_e c^2$. From there the system takes about $100 \, r_\mathrm{g} /c$ to reach a steady state.

We use a damping layer at the outer boundary to absorb all outgoing electromagnetic waves in order to mimic an open boundary. Particles are removed if $r\le r_\mathrm{h}$ or $r\ge r_\mathrm{max}$. We performed our runs with a grid resolution $1536 \, (r) \, \times \, 1024 \, (\theta)$, with the requirement that we resolve the plasma skin depth everywhere, which was checked {a posteriori}. The fiducial density of the simulations is the Goldreich-Julian density $n_\mathrm{GJ}=B_0 \omega_\mathrm{BH} / ( 4 \pi c e)$, which is needed to screen the electric field induced by the rotation of the black hole. In this equation, $\omega_\mathrm{BH} = c a / (2 r_\mathrm{h})$ is the black hole angular velocity.

\subsection{Magnetic configuration} \label{sec:mag_conf}



\begin{figure}[h!]
    \resizebox{\hsize}{!}{\includegraphics{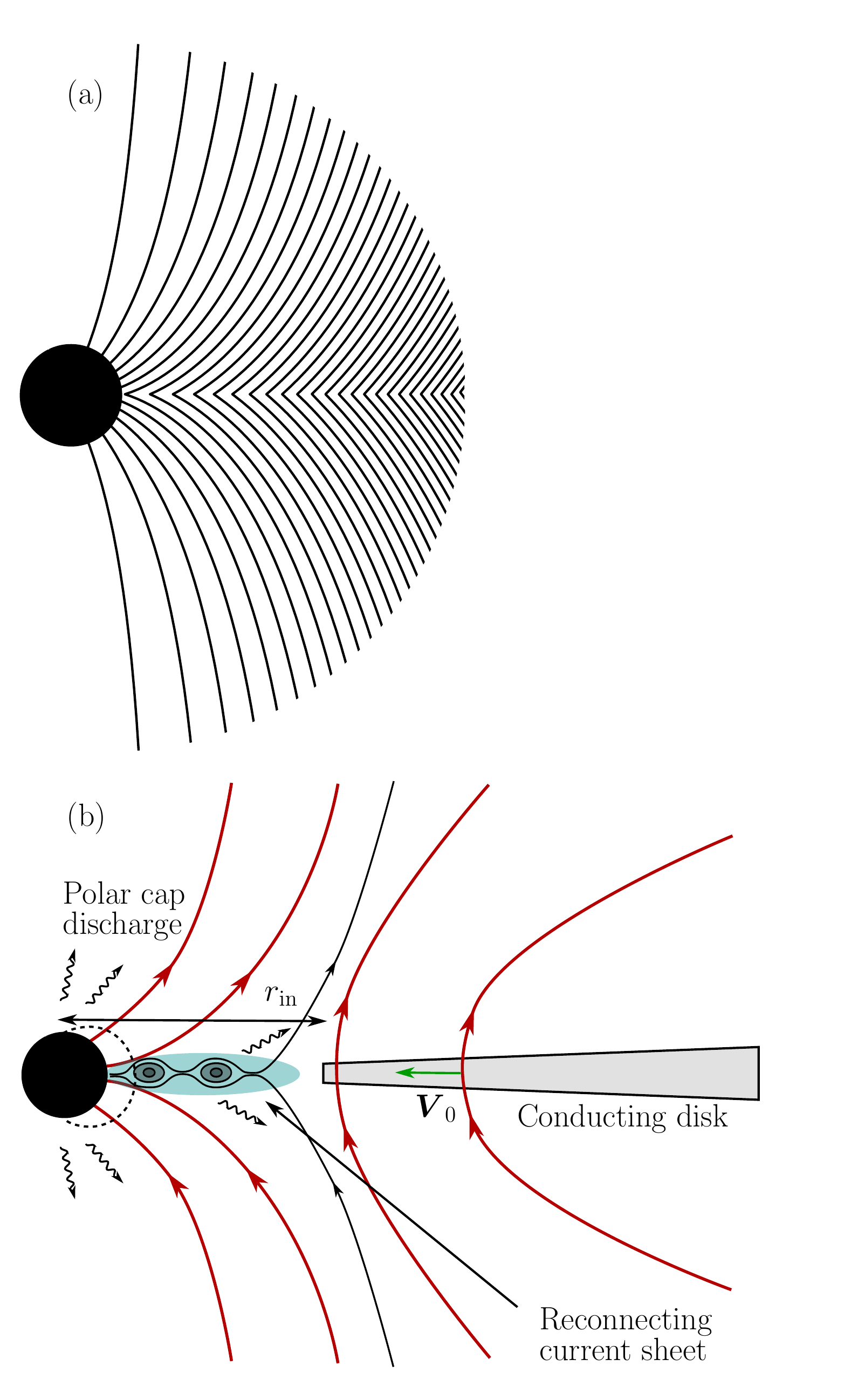}}
    \caption{Schematic magnetic configuration. (a) Initial poloidal magnetic field lines, according to Eq.~\eqref{eq:aphi}. (b) Magnetic configuration in steady state. Poloidal magnetic field lines are shown as red solid lines, except the last closed magnetic field line, which is in black. The equatorial current sheet (blue shaded area) is prone to the plasmoid instability. The conducting disk, represented by the gray shaded area, extends from $r_\mathrm{in}$ to the outer edge of the simulation box. Two emitting zones are highlighted: the polar cap (low inclination with respect to the spin axis) and the current sheet.}
    \label{fig:conducting_disk}
\end{figure}

We simulate a generic magnetic configuration with large-scale poloidal field. This choice is the natural setup of the BZ mechanism (but  see~\citealt{Parfrey_2015} and \citealt{Mahlmann_2020}), and it is suggested by GRAVITY observations of the Galactic center~\citep{Gravity_2018}. General relativistic magnetohydronamics (GRMHD) simulations of accretion flows also hint toward a paraboloidal geometry of the magnetic field lines~\citep{Komissarov_2007,McKinney_2012}. The initial poloidal magnetic field in the magnetosphere is defined for $\theta \leq \pi /2$ by the following flux function~\citep{Tchekhovskoy_2010}
\begin{equation} \label{eq:aphi}
    A_\varphi (r,\theta) = B_0 r_\mathrm{g}^2 \left( \dfrac{r+r_0}{r_\mathrm{h} +r_0} \right)^\nu ( 1 - \cos{\theta}),
\end{equation}
where $r_0$ and $\nu$ are free parameters. We chose $r_0=10 \, r_\mathrm{g}$ and $\nu=3$ in our runs. The geometry of the initial magnetic field lines is shown in the upper panel of Fig.~\ref{fig:conducting_disk}. In addition, in opposition to the work conducted in \citetalias{Crinquand_2020}, this magnetic configuration naturally produces anti-parallel field lines at the equator because they are dragged toward the black hole during the simulation. This allows an equatorial current sheet to develop due to the discontinuity of the magnetic field. However, it is not known whether a configuration with an equatorial current sheet extending beyond the ergosphere is stable. In the Wald configuration studied by~\citet{Parfrey_2019} the current sheet did not extend beyond the ergosphere. 

A magnetosphere with such an initial magnetic field quickly dies out after a few tens of $r_\mathrm{g}/c$ (see Sect.~\ref{sec:toy}). This occurs because    the current sheet extends to the outer boundary of the box, which is endowed with open boundary conditions. Magnetic reconnection at the current sheet ejects plasmoids and magnetic flux from the simulation box. Too much energy and flux are lost by the simulation box, and  the black hole almost completely expels the magnetic field lines threading it. Unlike pulsars, black holes do not have a conducting surface and cannot sustain a magnetic field. Therefore, we implemented a static and perfectly conducting disk as a boundary condition for the electromagnetic fields in the equatorial plane. This disk extends from its inner radius $r=r_\mathrm{in}$ to the outer boundary of the box, for $\theta \in [ \pi / 2 - \theta_0, \pi/2 + \theta_0]$, and we fixed $r_\mathrm{in}=6 \, r_\mathrm{g}$ and $\theta_0=0.02$ in all simulations. The resulting setup is represented the lower panel of Fig.~\ref{fig:conducting_disk}. Magnetic field lines crossing this disk are frozen in. The magnetic flux cannot escape the simulation box, which prevents our simulations from decaying entirely (see Fig.~\ref{fig:long_term}). We exclude the study of the magnetic linkage that can exist between the black hole and the disk, which is deferred to future work. We do not claim to simulate a realistic accretion disk, but rather to provide the physical conditions suitable to the study of the intrinsic behavior of the magnetosphere. The disk is merely included as a boundary condition for the fields. We are not interested in the zone surrounding the disk, and focus on the magnetosphere itself, that is, the zone enclosed by the field lines crossing the ergosphere. For this reason, in all subsequent figures we choose to leave the inner radius of the disk out of the represented domain.  We also checked that there was no significant numerical diffusion, and hence no unphysical slippage of the field lines.

\subsection{General features} \label{sec:results_dynamics}

We first describe the general features of our simulations before addressing the influence of magnetic field transport and the long-term evolution of the magnetosphere.  \\


\begin{figure}[t!]
    \resizebox{1.0\hsize}{!}{\includegraphics{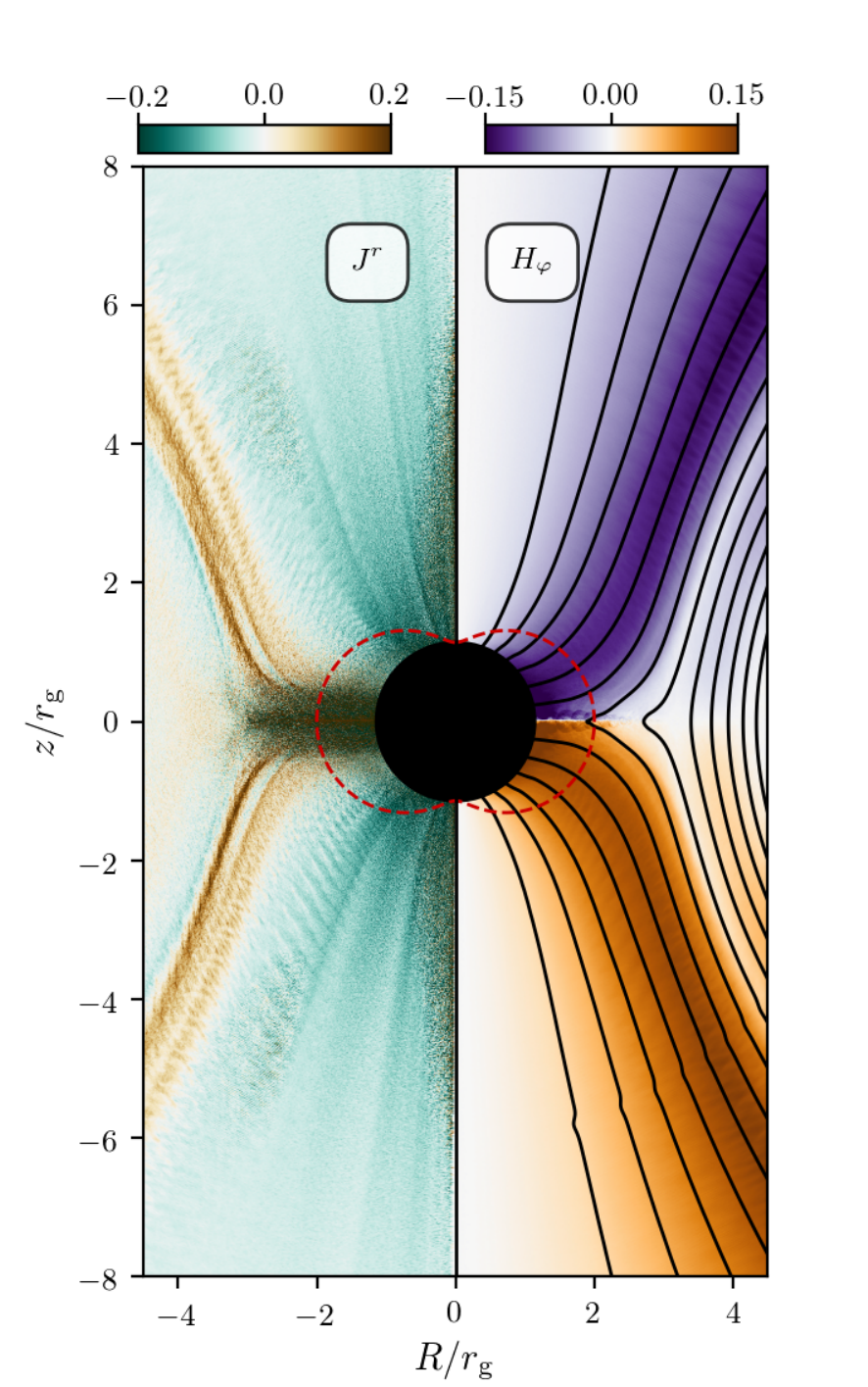}}
    \caption{Maps of the steady-state quantities. Left panel: Normalized electric current density $J^r / e c n_\mathrm{GJ}$. Right panel: Normalized toroidal component $H_\varphi / B_0 r_\mathrm{g}$. Poloidal magnetic field lines are represented by black solid lines. The dashed red line indicates the surface of the ergosphere. The quantities are averaged between $t=100 \, r_\mathrm{g}/c$ and $110 \, r_\mathrm{g}/c$.}
    \label{fig:structure}
\end{figure}

\begin{figure}[t!]
    \resizebox{1.0\hsize}{!}{\includegraphics{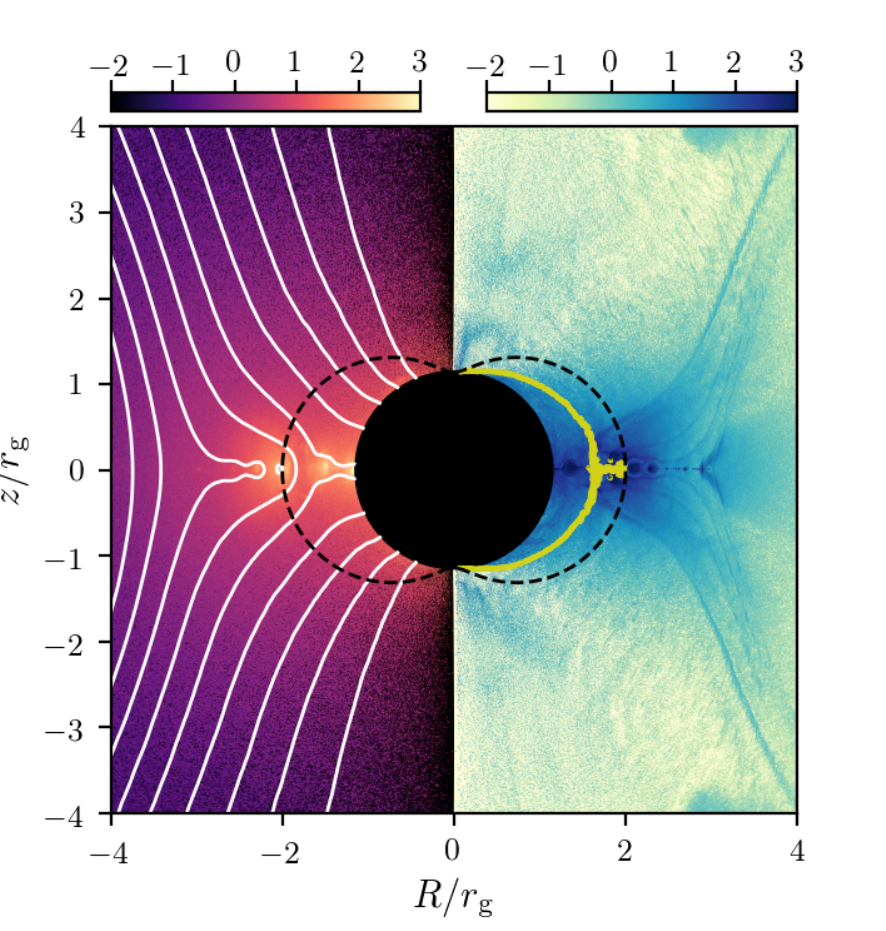}}
    \caption{Snapshot of the simulation with $\tau_0=50$ and $V_0 /c = 0.05$ at $t=126 r_\mathrm{g}/c$. Left panel: Logarithm of the normalized density of photons above the threshold, normalized by $n_\mathrm{GJ}$. Poloidal magnetic field lines are represented by white solid lines. Right panel: Logarithm of the normalized plasma density, normalized by $n_\mathrm{GJ}$. The dashed black line indicates the surface of the ergosphere, and the yellow solid line on the right panel shows the inner light surface. A movie is available online.}
    \label{fig:density}
\end{figure}

\begin{figure*}[!h]
    \includegraphics[width=17cm]{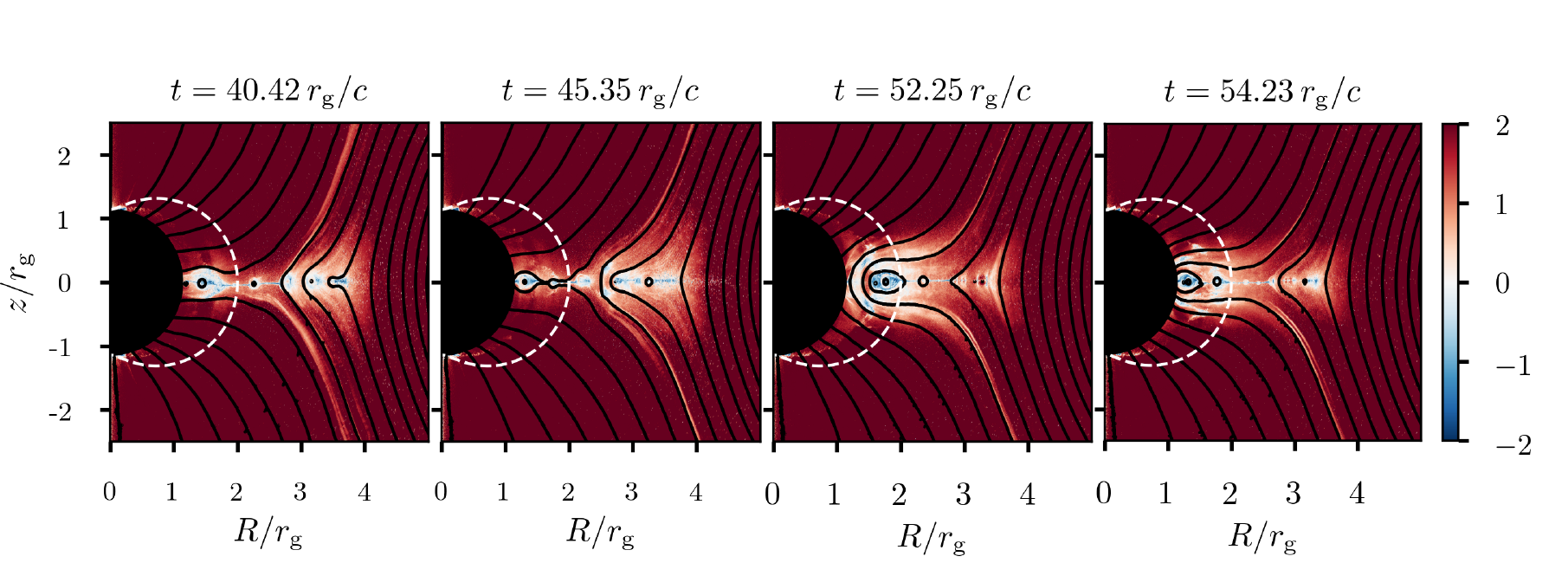}
    \caption{Snapshots of the logarithm of $\sigma = B^2 / 8 \pi n m_\mathrm{e} c^2 \Gamma$ from a simulation with $\tau_0=50$. The zones with $\sigma \simeq 1$ are in white. Poloidal magnetic field lines are represented by solid black  lines. The dashed white line indicates the surface of the ergosphere. The full temporal evolution is available as an online movie.}
    \label{fig:snapshots}
\end{figure*}

\para{Structure} The structure of the magnetosphere is shown in Fig.~\ref{fig:structure}. The right panel shows $H_\varphi$, which quantifies the poloidal current through a loop at constant $(r,\theta)$ according to Amp\`ere's law $\rot{\vec{H}}=4 \pi \vec{J} /c$. This toroidal field is nonzero on the field lines connected to the black hole, penetrating the ergosphere; therefore, a nonvanishing flux of energy and angular momentum can flow along those lines. The left panel shows the radial component of the current density. The electric current system is consistent with what is expected for a black hole magnetosphere in the force-free regime with $a>0$~\citep{Komissarov_2004a}. Our simulations have $\vec{\Omega} \cdot \vec{B} > 0$ in both hemispheres. In the upper hemisphere an electric field pointing toward the black hole is gravitationally induced by the frame-dragging of magnetic field lines. Negative poloidal currents are generated, which help screen the initial nonzero $\vec{D} \cdot \vec{B}$, thus giving rise to a negative $H_\varphi$. The situation is opposite in the lower hemisphere. By symmetry, $H_\varphi$ must vanish in the equatorial plane. The subsequent current sheet carries a positive electric current, closing the electric current system. This positive current flows along the separatrix, \textit{i.e.,} the last magnetic field line connected to the black hole, which defines the magnetospheric boundary. \\

\para{Pair creation} The equatorial current sheet is prone to  plasmoid instability~\citep{Uzdensky_2010}, which mediates fast magnetic reconnection. Magnetic energy is dissipated and deposited into particles, leading to intense pair creation. Figure~\ref{fig:density} shows a snapshot of the photon density above the pair-creation threshold and particle density, both in logarithmic scale.

We confirm that the mechanism described in \citetalias{Crinquand_2020} is still operating in this new configuration. Bursts of pair creation occur in an intermittent manner near the inner light surface\footnote{Light surfaces are the two surfaces separating subluminal and superluminal rotation for a point orbiting a Kerr black hole with angular velocity $\Omega$~\citep{Komissarov_2004a}. The relevant inner light surface is defined by taking $\Omega$ as the velocity of the magnetic field lines. It is located within the ergosphere.} at intermediate latitudes. This fresh plasma mostly follows the magnetic field lines; therefore, it mainly flows close to the magnetospheric boundary. Inside the bursts the plasma density is marginally denser than the Goldreich-Julian density, and the outflowing plasma is highly magnetized. Pair creation is almost quenched near the rotation axis. We checked that in this zone the $4$-current is null, although it is spacelike near the horizon at intermediate latitudes. In addition, the acceleration of particles in the X-points of the current sheet triggers pair creation and high-energy photon emission. The plasma density can reach $10^3 n_\mathrm{GJ}$ in the current sheet plasmoids. \\

\para{Dynamics} The magnetosphere displays an interesting dynamical phenomenon that is  responsible for the replenishment of the magnetic field threading the black hole (see Fig.~\ref{fig:snapshots}). Starting from an initial state similar to that shown in Fig.~\ref{fig:structure}, plasma accumulates near the Y-point of the magnetosphere\footnote{The Y-point is defined as the point of the current sheet most distant to the black hole. It is the point where the two separatrices diverge from the equator.}. This plasma is supported by the magnetic pressure inside the magnetosphere. When the magnetosphere can no longer sustain the plasma, which roughly occurs when the particle energy density exceeds the magnetic energy density, a giant plasmoid forms and suddenly plunges into the black hole. This corresponds to the breakdown of the force-free approximation. The weakly magnetized plasma plunges due to the gravitational pull of the black hole, and works against the magnetic tension of field lines. As this giant plasmoid rushes inward, it pulls inward vertical magnetic field lines that were not crossing the event horizon initially. This replenishes the magnetic flux of the black hole. After the black hole swallows the giant plasmoid, the magnetosphere goes back to its initial state, until a new giant plasmoid is formed.

\subsection{Long-term evolution}

\begin{figure*}[!h]
    \includegraphics[width=17cm]{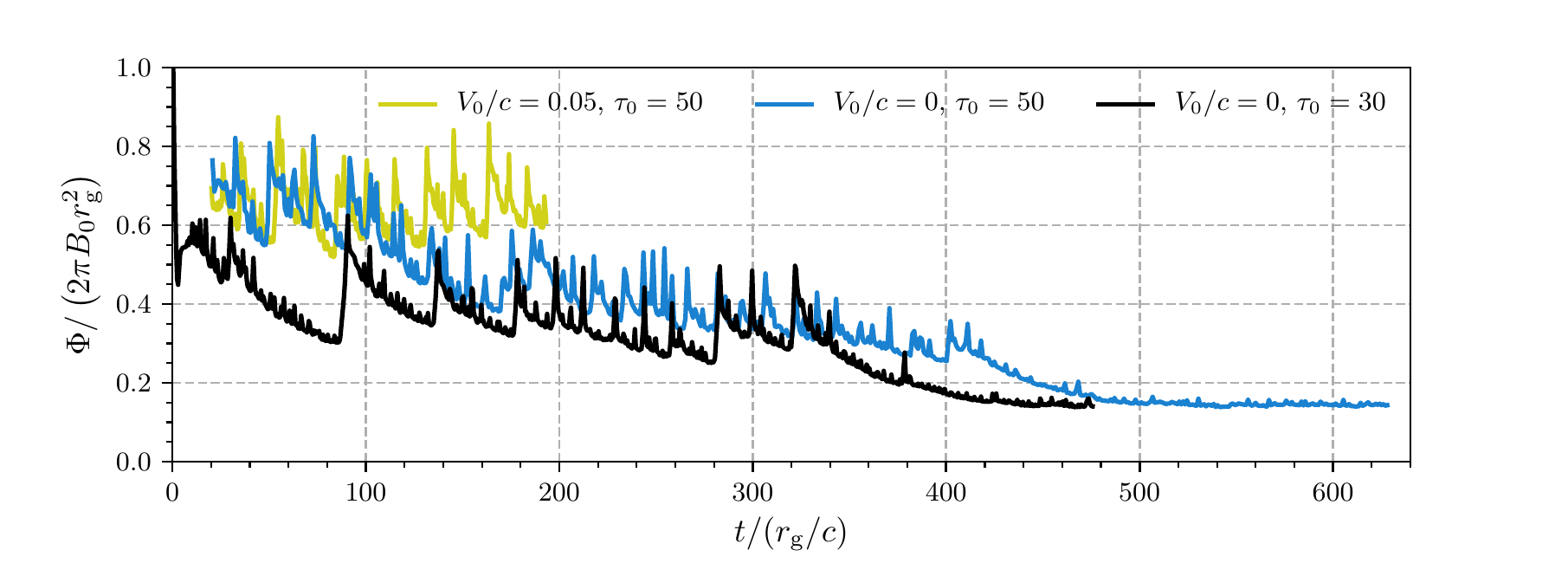}
    \caption{Evolution of the magnetic flux $\Phi / 2 \pi B_0 r_\mathrm{g}^2$ through the upper hemisphere of the event horizon for different $V_0$ and $\tau_0$.}
    \label{fig:long_term}
\end{figure*}

\begin{figure}[t!]
    \resizebox{1.0\hsize}{!}{\includegraphics{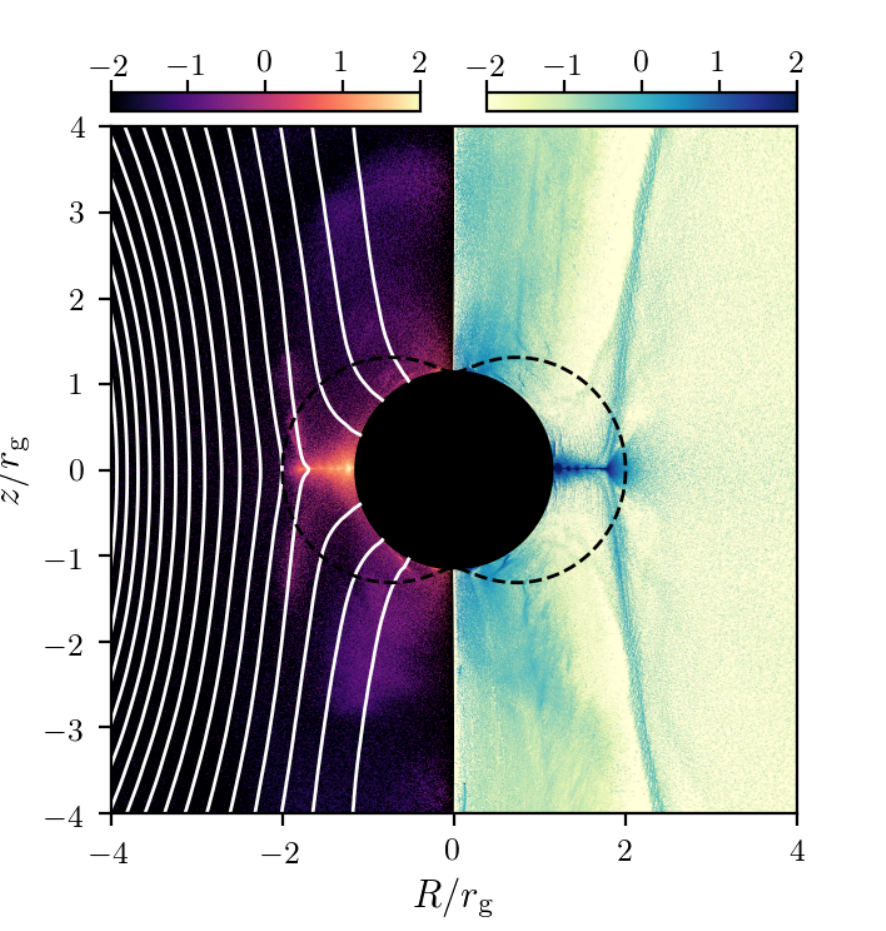}}
    \caption{Snapshot of the simulation with $\tau_0=50$ and $V_0=0$ at $t=611 r_\mathrm{g}/c$. Left panel: Logarithm of the normalized photon density, normalized by $n_\mathrm{GJ}$. Poloidal magnetic field lines are represented by white solid lines. Right panel: Logarithm of the normalized plasma density, normalized by $n_\mathrm{GJ}$. The dashed black line indicates the surface of the ergosphere.}
    \label{fig:density_v_0}
\end{figure}

The outcome of the simulation is shown by the black and blue curves in Fig.~\ref{fig:long_term}, which represent the evolution of the magnetic flux $\Phi$ through the upper hemisphere of the event horizon with time. The magnetosphere experiences the dynamic cycles described in the previous section for about $300 \, r_\mathrm{g}/c$, but the magnetic flux $\Phi$ decays secularly. It settles at a steady value after a time $\simeq 500 \, r_\mathrm{g}/c$. The steady state of the simulation resembles the Wald setup~\citep{Parfrey_2019}, as can be seen in Fig.~\ref{fig:density_v_0}. The field lines are much more vertical and, more importantly, the Y-point is located very close to the boundary of the ergosphere. In this steady state there are no more giant plasmoid accretion cycles. The current sheet is still disrupted by the tearing instability, so that small plasmoids fall toward the black hole. The magnetic flux escape by magnetic reconnection is exactly balanced by the supply of magnetic flux caused by the inflowing plasmoids pulling vertical field lines. Therefore, without external forcing, the only stable configuration for the magnetosphere is Wald-like. This is reminiscent of pulsar magnetospheres where the Y-point naturally migrates toward an equilibrium position at the light cylinder~\citep{Spitkovsky_2006}. This configuration is close to force-free, except in the current sheet. 

It should be   noted  that because the magnetic field strength has dropped significantly, the maximum Lorentz factor that particles can achieve is no longer much larger than the pair-creation threshold. Being slightly starved, large gaps can momentarily open up. This is merely an effect of our limited scale separation and does not affect the general conclusion. We also note that the final value for the magnetic flux does not depend on $\tau_0$ in the range of parameters we have tested. If the opacity of the medium is high enough, the magnetosphere can reach a state close to force-free, irrespective of the plasma supply details.

\subsection{Magnetic field transport}

We are interested in maintaining the dynamic state and impeding magnetic field decay, since this variable state is promising for the prospect of high-energy flares. Therefore, we added the possibility of supplying magnetic flux to the central black hole in order to study the response of the magnetosphere either free or forced. To this end, we did not inject magnetic flux in the whole simulation box, but rather advected the frozen-in field lines that are initially crossing the perfectly conducting disk. We added a small toroidal electric field $\vec{E}_\mathrm{acc}=- \left( \vec{V}_0 /c \right) \times \vec{B}$ only in the conducting disk. We ran another set of simulations of varying $\tau_0$, but this time  with $V_0 / c = 0.05$. This setup can mimic inward magnetic flux transport in accretion flows~\citep{Lubow_1994}. The latter value of $V_0$ that we used is consistent with ideal MHD simulations of accretion disks~\citep{Jacquemin_2020}.

By virtue of Faraday's law applied to a loop of radius $r=r_\mathrm{in}$ in the equatorial plane, the magnetic flux through a surface enclosed by this loop must increase steadily for $V_0 \neq 0$ and remain constant for $V_0=0$. In other words, magnetic field lines that have been transported below $r=r_\mathrm{in}$ at $\theta=\pi/2$ must remain below $r_\mathrm{in}$ from then on. This is why we place the inner boundary of the conducting disk at sufficient distance $r_\mathrm{in}=6 \, r_\mathrm{g}$ from the black hole. We choose not to run simulations with $V_0 \neq 0$ for as long as simulations with $V_0=0$ because the magnetic flux within $r=r_\mathrm{in}$ would accumulate near the ergosphere.

\begin{figure}[h!]
    \resizebox{\hsize}{!}{\includegraphics{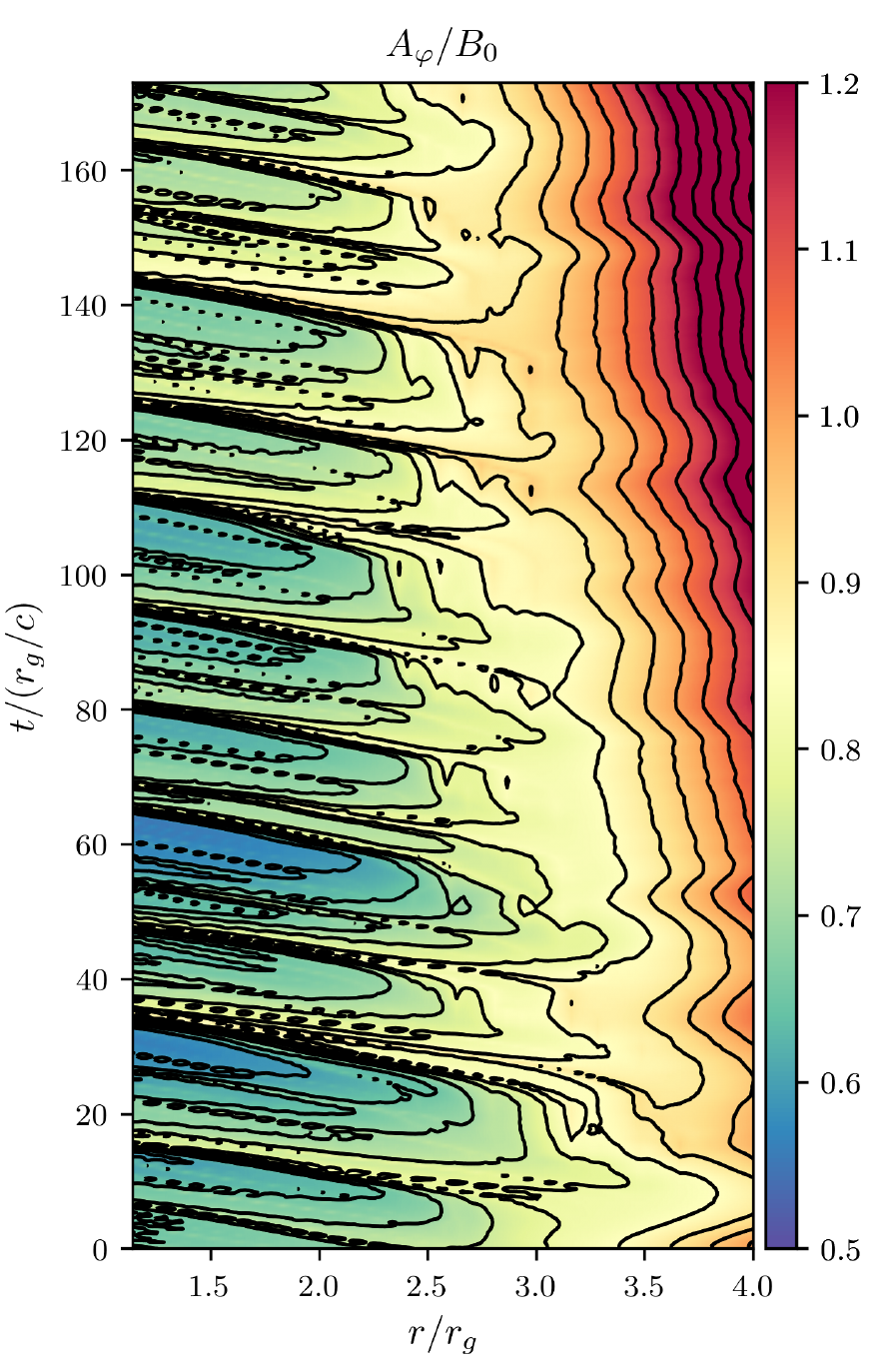}}
    \caption{Spacetime diagram of $A_\varphi$ in the equatorial plane for the simulation with $\tau_0=50$ and $V_0/c = 0.05$. The black solid lines are the contours of $A_\varphi$, which represent poloidal magnetic field lines.}
    \label{fig:aphi}
\end{figure}

\begin{figure*}[t!]
    \includegraphics[width=17cm]{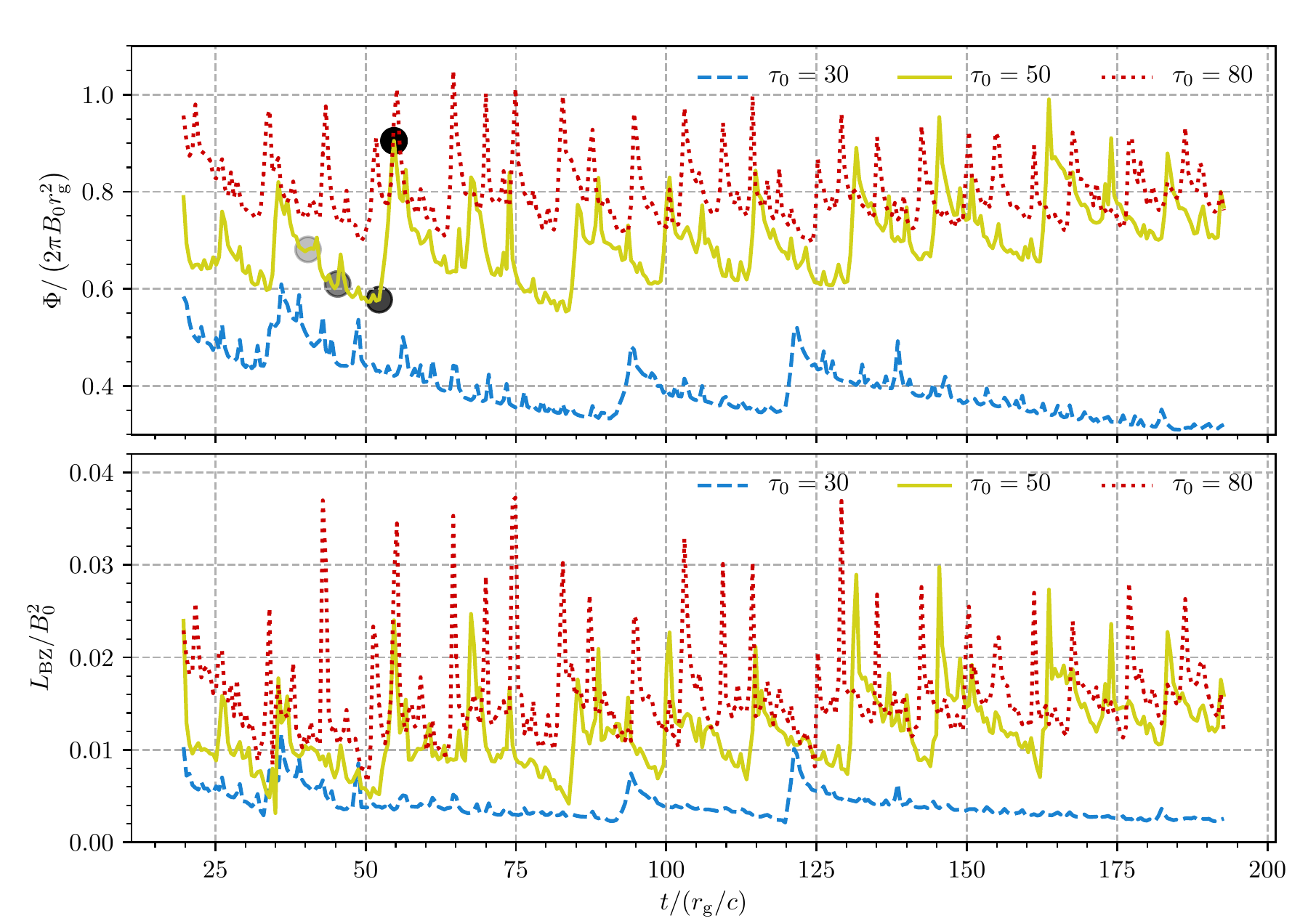}
    \caption{Time evolution of the normalized magnetic flux $\Phi / 2 \pi B_0 r_\mathrm{g}^2$ through the upper hemisphere of the event horizon (upper panel) and of the normalized Poynting flux $L_\mathrm{BZ} / B_0^2$, integrated over the event horizon (lower panel), for the three simulations at $V_0/c=0.05$. The black disks, relative to the $\tau_0=50$ simulation, indicate the times of the snapshots relative to Fig.~\ref{fig:snapshots}.}
    \label{fig:phi}
\end{figure*}

In the presence of magnetic field line transport ($V_0 \neq 0$) the magnetosphere is able to remain in a dynamic state of periodic giant plasmoid accretion events (Fig.~\ref{fig:long_term}), and the inflow of magnetic flux compresses the magnetosphere and compensates the secular decay. The evolution of $\Phi$ in this dynamic state is represented in the upper panel in Fig.~\ref{fig:phi} for different optical depths, with the four blue dots representing successive snapshots relative to Fig.~\ref{fig:snapshots}. As a magnetized giant plasmoid is swallowed by the black hole, the magnetic flux experiences a sharp rise. Between two successive giant plasmoid accretion events the magnetic flux decays almost exponentially with time due to magnetic reconnection. We observed that the characteristic decay time of $\Phi$ barely depends on $\tau_0$. On the other hand, the frequency of these accretion cycles is controlled by the fiducial optical depth $\tau_0$: it increases with increasing $\tau_0$. This occurs because mass loading at the Y-point is more efficient at high optical depth, which results in more frequent cycles of accretion. These cycles are illustrated in Fig.~\ref{fig:aphi}, which represents a spacetime diagram of the flux function $A_\varphi$ in the equatorial plane. They occur with a period of around $15 \, r_\mathrm{g} /c$. The slow transport of magnetic field lines from the conducting disk to the black hole is also visible between $3 \, r_\mathrm{g}$ and $4 \, r_\mathrm{g}$.

The lower panel in Fig.~\ref{fig:phi} shows the time evolution of the Poynting flux through the event horizon for the simulations with magnetic field transport. It is defined as
\begin{equation}
    L_\mathrm{BZ} = \iint_\mathcal{B} S^r \sqrt{h} \diff{\theta} \diff{\varphi},
\end{equation}
where $h$ denotes the determinant of the spatial $3$-metric, the integration is performed over the event horizon $\mathcal{B}$ at $r=r_\mathrm{h}$, and

\begin{equation}
    S^r = \dfrac{1}{4 \pi} \dfrac{1}{\sqrt{h}} \left( E_\theta H_\varphi - E_\varphi H_\theta \right)
\end{equation}
is the radial component of the Poynting vector, \textit{i.e.,} the flux of electromagnetic energy-at-infinity~\citep{Komissarov_2004a}. We also observe sharp rises in the Poynting flux, synchronized with those in $\Phi$. This comes as no surprise since the output power is expected to scale as $\Phi^2$ if the Blandford-Znajek process is activated \citep{Blandford_1977,Tchekhovskoy_2010}. In the case of a pure split-monopole magnetosphere the total Poynting luminosity is $L_0 = B_0^2 \omega_\mathrm{BH}^2 /6$. The measured luminosity is lower than this estimate because some flux is removed from the event horizon during an initial transient, due to the initial conditions not being an equilibrium state. This also explains why $\Phi$ is consistently below $2 \pi r^2_\mathrm{g} B_0$.

The time-averaged luminosity corresponds to the total energy extracted from the black hole by the Blandford-Znajek process. In Sect.~\ref{sec2} all energy fluxes and light curves are   normalized by this average luminosity $\langle L_\mathrm{BZ} \rangle$.

\subsection{Toy model for magnetic flux decay} \label{sec:toy}

The decay of the magnetic flux $\Phi$ through the event horizon, in the absence of any source term due to inflowing plasmoids, is a consequence of dissipation of magnetic energy by magnetic reconnection. We  provide a toy model to account for the order of magnitude of the characteristic time $T$, assuming axisymmetry. The magnetic flux $\Phi$ can be expressed as

\begin{equation}
    \Phi = \iint_{\mathcal{B}_+} B^r \sqrt{h} \diff{\theta} \diff{\varphi} = 2 \pi A_\varphi (r_\mathrm{h}, \pi /2),
\end{equation}
where $\mathcal{B}_+$ is the upper hemisphere of the event horizon. Faraday's law allows us to express the time derivative of $\Phi$ as the circulation of $\vec{E}$ along a loop of radius $r_\mathrm{h}$ in the equatorial plane:
\begin{align}
    \ddroit{\Phi}{t} = 2 \pi \partial_t A_\varphi (r_\mathrm{h}, \pi /2) = - 2\pi c E_\varphi (r_\mathrm{h}, \pi /2).
\end{align}
In a purely axisymmetric and stationary magnetosphere, we have $E_\varphi=0$ everywhere. However, this component arises in and near the current sheet because there is an inflow of plasma at velocity $\vec{V}_\mathrm{in} = V_\mathrm{in} \partial_\theta$ toward the reconnection region. Just above and below the current sheet, the electric field reads $\vec{E} = - \left( \vec{V}_\mathrm{in} /c \right)\times \vec{B}$. At such high magnetizations, the outflow velocity is very close to $c$. We then define a global dimensionless reconnection rate $R$ as
\begin{equation} \label{eq:reconnection_rate}
    R = \dfrac{\abs{V_\mathrm{in}}}{c} = r_\mathrm{g} \dfrac{E_\varphi}{\sqrt{h} \abs{B^r}},
\end{equation}
where $h$ is evaluated at $r=r_\mathrm{h}$ and $\theta= \pi /2$. Here $V_\mathrm{in}$ is not measured by the FIDO, but with respect to the grid. We also assume that the configuration of field lines at the event horizon is close to a split monopole, so that $B^r (r=r_\mathrm{h})$ does not depend much on $\theta$. The magnetic flux can be written as $\Phi = B^r (r_\mathrm{h}) \mathcal{S}$, where $\mathcal{S} = 2 \pi \left( r_h^2 + (a r_\mathrm{g} )^2 \right)$ is half the area of the event horizon~\citep{Bicak_1985}. Ultimately, the evolution of the magnetic flux is governed by

\begin{equation}
    \ddroit{\Phi}{t} = - \dfrac{2 \pi \sqrt{h}}{\mathcal{S}} \dfrac{c}{r_\mathrm{g}} R \, \Phi.
\end{equation}
If the global reconnection rate is time-independent, the magnetic flux decreases exponentially with a characteristic decay time 
\begin{equation}
    T = \dfrac{\mathcal{S}}{2 \pi \sqrt{h}} \dfrac{1}{R} \dfrac{r_\mathrm{g}}{c} \simeq \dfrac{1}{R} \dfrac{r_\mathrm{g}}{c}
.\end{equation}

From Figs.~\ref{fig:phi} and \ref{fig:long_term} we measure the slope of the exponential decay, and obtain a reconnection rate $R = 0.02 \pm  0.002$, corresponding to a decay time $T \simeq 50 \, r_\mathrm{g}/c$. We also measured the local reconnection rate using Eq.~\eqref{eq:reconnection_rate} and found values ranging from $0.02$ to $0.04$, which is consistent with the flux decay time. This model naturally explains why the characteristic decay time is the same in all simulations. 

The local reconnection rate in collisionless relativistic reconnection has been determined by numerous numerical studies \citep[e.g.,][]{Werner_2018}, with typical values between $10^{-2}$ and $10^{-1}$. To compare with the decay time $T$, as measured by an observer at infinity, one needs to take into account gravitational dilation. Assuming the current sheet is roughly comoving (radially inward) with the Kerr-Schild FIDO at $r=r_\mathrm{h}$ and $\theta=\pi/2$, by definition of the lapse function $\alpha$, the local reconnection rate can be estimated as $R/\alpha \simeq 1.66 \, R$. Our value of $R$ is consistent with measurements of the local collisionless reconnection rate, although slightly lower. 



As mentioned in Sect.~\ref{sec:mag_conf}, we also ran simulations with no conducting disk. In these simulations the equatorial current sheet quickly extends across the whole box. The initial magnetic energy is quickly dissipated, whereas the mechanism previously analyzed cannot take place; there is no available vertical magnetic flux that could compensate for this decay. The simulation is exhausted of particles and dies out after a time on the order of $T$, which is solely determined by the reconnection rate.

\section{Synthetic light curves} \label{sec2}

\subsection{Numerical method}

\begin{figure*}[h!]
    \includegraphics[width=17cm]{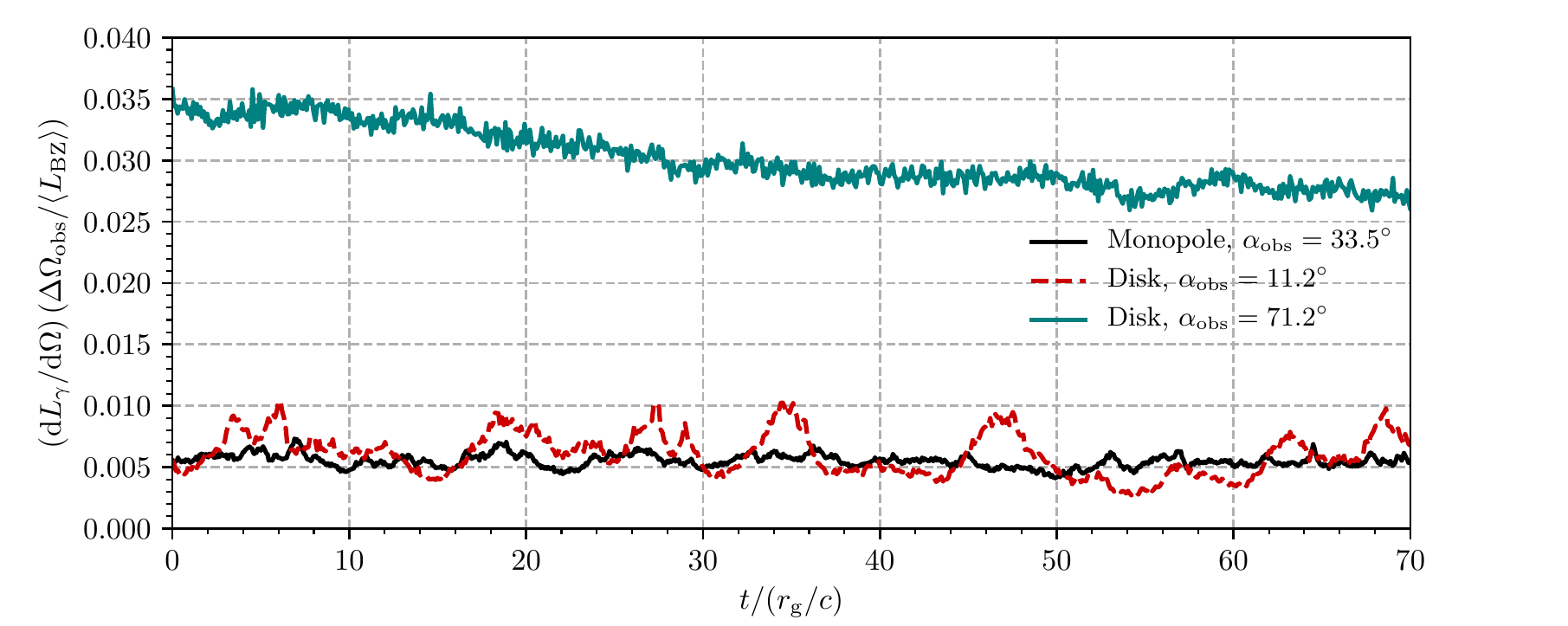}
    \caption{Light curve for the monopole simulation at $\tau_0=30$ (described in \citetalias{Crinquand_2020}), normalized by $L_0= B_0^2 \omega_\mathrm{BH}^2 /6$, at $\alpha_\mathrm{obs} = 33.5^\circ$ (black solid line), and light curves for the disk simulation with $\tau_0=50$ and $V_0/c = 0.05$ at two different viewing angles: $\alpha_\mathrm{obs} = 11.2^\circ$ (red dashed line) and $\alpha_\mathrm{obs} = 71.2^\circ$ (blue solid line). These two  light curves are normalized by the time-averaged value of $L_\mathrm{BZ}$ shown in Fig.~\ref{fig:phi}.}
    \label{fig:light curves}
\end{figure*}


As discussed in Sect.~\ref{sec:techniques}, we must save the information carried by photons that are below the pair-creation threshold. Given their initial positions, directions, and emission times, our goal is to reconstruct light curves for different viewing angles (with respect to the spin axis). This task has already been performed in flat spacetime~\citep{Cerutti_2016} in the context of pulsar magnetospheres. Here this approach must be generalized to photons propagating in a curved spacetime. We neglect the gravitational influence of the black hole beyond a given $r_\mathrm{out}$, which we fix at $200 \, r_\mathrm{g}$: for $r \ge r_\mathrm{out}$, photons are considered to propagate in straight lines. We need to integrate the null geodesics of the photons from their emission points up to $r=r_\mathrm{out}$ in order to compute their final directions and times of flight. Then, just as in flat spacetime, the light curve can be reconstructed if the directions of the outgoing photons are known at $r=r_\mathrm{out}$.

Keeping these photons in the simulation box and integrating their equations of motion with the PIC algorithm, even with a looser constraint on the time step, would be too demanding computationally. As it happens there is no need to solve the entire geodesic since the only relevant information is the initial and final coordinates $(t,r,\theta,\varphi)$ of the photons. Instead, we use the public ray-tracing code $\texttt{geokerr}$~\citep{Dexter_2009} . This code is optimized to integrate numerically null geodesics in the Kerr metric, and allows us to directly compute the final coordinates. If a photon produced by IC scattering is measured to be under the pair-creation threshold, its relevant information is dumped to a file that will be processed by \texttt{geokerr} before being discarded. All diagnostics can then be performed in post-processing. The light curve reconstruction procedure and the coupling between the two codes are detailed in Appendix~\ref{appendix}.


\subsection{Results}

We applied the previously outlined procedure to our three simulations with $V_0/c=0.05$, along with the highest opacity monopole simulation presented in \citetalias{Crinquand_2020} (which had $\tilde{B}_0 = 5 \times 10^5$, $\tilde{\varepsilon}_0=5 \times 10^{-3}$ and $\tau_0=30$). The time resolution is $\Delta t = 0.0098 r_\mathrm{g} /c$, and the angular resolution is $\Delta \alpha_\mathrm{obs} = \pi / 24$. We compute the energy flux per unit of solid angle. Some resulting light curves, normalized by the time-averaged BZ power of each simulation $\langle L_\mathrm{BZ} \rangle$ and by the solid angle $\Delta \Omega_\mathrm{obs} = 2 \pi \sin{\alpha_\mathrm{obs}} \Delta \alpha_\mathrm{obs}$, are shown in Fig.~\ref{fig:light curves}. The monopole light curves do not depend much on the viewing angle, especially at intermediate latitudes. One example is shown in Fig.~\ref{fig:light curves}. This is expected since the monopole simulations show little structure in the orthoradial direction, and photons are mainly emitted radially by particles flowing along the magnetic field lines. We find a bolometric luminosity of $\langle L_\gamma \rangle = \langle \int \left( \diff{L_\gamma} / \diff{\Omega} \right) \diff{\Omega} \rangle \simeq 0.04 \, L_0 \simeq 0.04 \, \langle L_\mathrm{BZ} \rangle$. This is consistent with the dissipation rate of electromagnetic energy measured in \citetalias{Crinquand_2020}, and confirms the hypothesis that the dissipated Poynting flux is mainly transferred to photons below the pair-creation threshold. Although the light curve shows signs of rapid variability, which is consistent with the small size of the gap, no flare can be detected. The incoherent process of pair creation along various magnetic field lines hinders the occurrence of large amplitude flares.

The situation is rather different for the disk simulations with external forcing (Fig.~\ref{fig:light curves}). These light curves show pronounced differences if viewed  face-on (line of sight close to the spin axis, low $\alpha_\mathrm{obs}$) or edge-on (line of sight close to the equatorial plane, $\alpha_\mathrm{obs} \simeq \pi /2$). At low $\alpha_\mathrm{obs}$ they exhibit strong variability. During a flaring event the flux doubles within a rising time $\simeq 2 \, r_\mathrm{g} /c$. The periodicity of these flares is around $10 \, r_\mathrm{g}/c$, in agreement with the periodicity of the giant plasmoid accretion cycles. Conversely, light curves observed at $\alpha_\mathrm{obs} \simeq \pi /2$ are remarkably smooth, with no sign of variability at all. In order to understand these qualitative differences, we  constructed two light curves, associated with the sites of emission of the photons. We  distinguished the polar cap as the zone defined by $\theta \in [0^\circ , 60^\circ] \cup [120^\circ , 180^\circ]$, and the current sheet, defined by $\theta \in [60^\circ , 120^\circ]$ and $r\in [r_\mathrm{h}, r_\mathrm{in}]$. Unsurprisingly, photons emitted in the polar cap mainly contribute to the emission at low viewing angles, whereas the emission at $\alpha_\mathrm{obs} \simeq \pi /2$ is mainly due to the current sheet.

In the simulations with external forcing the average bolometric luminosity $\langle L_\gamma \rangle$ ranges between $0.3 \, \langle L_\mathrm{BZ} \rangle$ and $0.5 \, \langle L_\mathrm{BZ} \rangle$. Therefore, a very significant fraction of the BZ power is converted to IC luminosity. The radiative efficiency is much higher than in the monopole simulations. The total luminosity of photons emitted in the polar cap is around $5\%$ of the BZ luminosity, a fraction that is similar to the monopole high-energy bolometric luminosity, although the polar cap emission is much more variable in these simulations. The current sheet and equatorial plasmoids emit 30\% to 40\% of the luminosity.

\begin{figure}[t!]
    \resizebox{1.0\hsize}{!}{\includegraphics{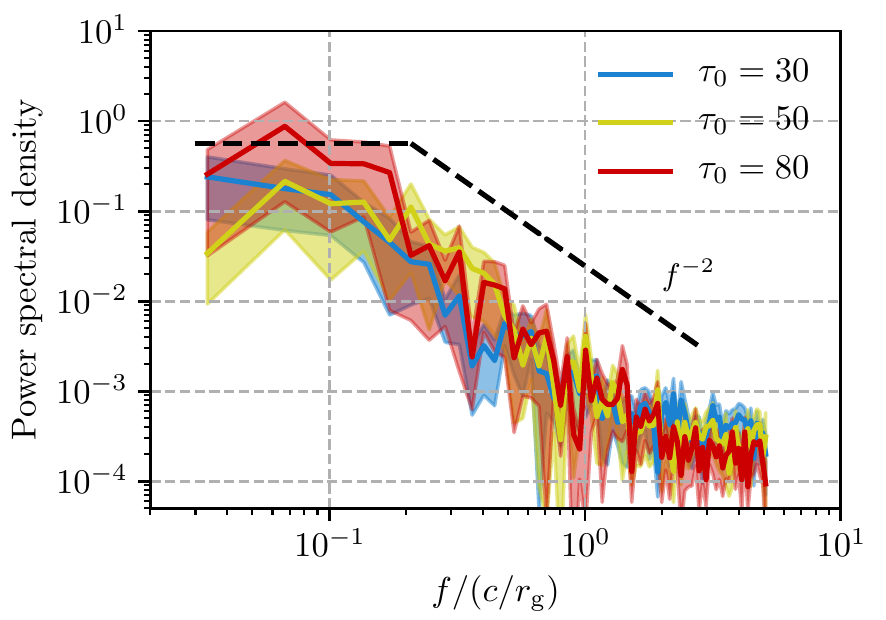}}
    \caption{Power spectral density of the light curves at viewing angle $\alpha_\mathrm{obs} =11.2^\circ$, for the simulations with $V_0/c=0.05$, as a function of the frequency $f$. The black dashed line shows the red-noise scaling $f^{-2}$, down to a spectral break located close to $0.1 \, c/r_\mathrm{g}$. The shaded areas indicate the error bars on the power spectra.}
    \label{fig:spectrum}
\end{figure}

High variability should not be expected from magnetospheres observed edge-on. This stems from the fact that the formation of plasmoids in the current sheet, and the subsequent emission of high-energy photons, is inherently incoherent. Furthermore, photons emitted in this region travel along complex null geodesics that can have several turning points in $\theta$. These geodesics are likely to differ significantly from simple radial rays; this adds to the decoherence that impacts current sheet emission, and erases any strong variability. On the other hand, photons emitted from the polar cap follow more direct geodesics toward the observer at infinity, such that the variability of the primary process is imprinted in the light curve. Therefore, the polar cap  shows more pronounced variability in these simulations than in the monopole case. This indicates that the gap dynamics cannot be studied with no consideration for the global magnetospheric structure: the magnetospheric dynamics enhance the activity of the gap.


The power spectral densities of the light curves at $\alpha_\mathrm{obs} = 11.2^\circ$ for the simulations with $V_0/c = 0.05$, computed using the \texttt{stingray} package \citep{Huppenkothen_2019}, are shown in Fig.~\ref{fig:spectrum} (after logarithmic rebinning). At frequencies $0.1 \, c/r_\mathrm{g} \lesssim f \lesssim 2 \, c/r_\mathrm{g}$, it is nicely described by a red-noise power law $\propto f^{-p}$, with $p= 2.00 \pm 0.13$. A spectral break is visible around $0.1 \, c/r_\mathrm{g}$. The spectral break frequency is consistent with the characteristic timescale associated with giant plasmoid accretion events. Resolving this spectral break observationally would require data acquisition for much longer than $10 \, r_\mathrm{g}/c$ (more than $4$ days in the case of M87*). Beyond $2 \, c/r_\mathrm{g}$, the power spectra are  similar to  white noise. Most of the power is distributed at lower frequencies: flux variations on long timescales dominate those on short timescales. The value of the index $p$ is in agreement with that measured by \citet{Aharonian_2007} from the AGN PKS 2155-304, although this measure may depend on whether the AGN is in a flaring state or not~\citep{Hess_2017}. We found that the value of $p$ does not depend much on $\tau_0$. The characteristic value of the plasma frequency $\nu_\mathrm{p} = \sqrt{8 \pi e^2 \langle n / \Gamma \rangle / m_\mathrm{e}} / 2 \pi$ is $50 \, c / r_\mathrm{g}$, and lies beyond the frequency range presented on the spectrum.

\section{Discussion}

We acknowledge that the results described in Sect.~\ref{sec1} would slightly differ in 3D since nonaxisymmetric modes would also allow the interchange of tenuous magnetospheric plasma with dense unmagnetized plasma at the Y-point. In particular, it is possible that $\Phi$ would not experience sharp and periodic peaks. Nonetheless, we believe this mechanism should still hold and allow the black hole to retain a significant magnetic flux and luminosity on a timescale longer than the characteristic reconnection decay time $T$.

In this paper we were   interested in the pure magnetospheric response of a low-luminosity AGN. We find that a substantial fraction of the BZ electromagnetic luminosity is channeled into IC photons, especially if magnetic flux is supplied to the black hole by the accretion flow. This high radiative efficiency is the most salient feature of our simulations, especially in the presence of external forcing. We note that emission from equatorial latitudes is smoothed out, such that high-energy variability should primarily be expected from the polar caps. Variability from the polar caps is enhanced with respect to the monopole simulations. However, even with constant external forcing, the intrinsic activity of a steady-state black hole magnetosphere does not reproduce the most dramatic features of AGN flares: a flux-doubling time below $r_\mathrm{g}/c$ and an increase in the flux by a factor of at least $5$, rather than $2$ in our modeling. The variability seems to be well characterized by a red-noise power law down to $\simeq 2 \, c/r_\mathrm{g}$. Because of the nonaxisymmetric modes mentioned earlier, 3D simulations are likely to show even less variability. It should also be noted that in our simulations, the background radiation field is mono-energetic. Using a more realistic power law would have reduced the variability in the gap screening because the pair-creation threshold would not have been defined at a single photon energy. This makes it even less likely that realistic gamma-ray flares can be accommodated with our numerical setup.


We are able to quantify how fast the magnetic flux through the upper hemisphere of the event horizon decays, and find that external forcing is necessary in order to sustain a dynamic magnetospheric state. A free magnetosphere naturally tends toward a steady state similar to the Wald configuration. Recently, \citet{Ripperda_2020} considered the possibility of magnetic reconnection powering infrared and X-ray flares in the Galactic center, and studied this scenario by resistive GRMHD simulations. They found that only in the magnetically arrested disk setup could there be a flaring state, during which plasmoids formed in an equatorial current sheet are heated to relativistic temperatures. This is consistent with our finding that in the absence of magnetic flux supply, the magnetosphere reaches a somewhat quiescent state and cannot produce flares. 

Although the numerical setup used here can seem idealized, we note that the magnetic configuration that we have studied is actually the natural outcome, at the ergospheric scale, of any larger-scale configuration of an isolated magnetosphere~\citep{Komissarov_2007}. Therefore, we think that our findings concerning the generic magnetospheric response have broad applicability. If of magnetospheric origin, rather being a manifestation of the intrinsic variability due to the pair production mechanism, flares could be interpreted as the fast response of a black hole magnetosphere to a sudden change in the external parameters. This conclusion was also reached by \citet{Levinson_2018} and \citet{Kisaka_2020} through radiative 1D GRPIC simulations. For example, a variation in the accretion rate would cause the density of soft photons to increase, leading to an increase in $\tau_0$. The velocity of magnetic field transport could also change, if a very magnetized plasma blob should accrete toward the magnetosphere. In that sense, black hole magnetospheres differ fundamentally from pulsar magnetospheres: pulsar activity is determined by parameters that are characteristic to the pulsar itself, and therefore remains quite stable. Another possibility is that flares result from a magnetosphere--disk interaction. This is suggested by the GRAVITY observation of a hot spot orbiting near the innermost circular orbit around Sgr A*~\citep{Gravity_2018}.

\begin{acknowledgements}
 
We thank Geoffroy Lesur and Amir Levinson for fruitful discussions that helped improve the article, and the anonymous referee for constructive and insightful suggestions. This project has received funding from the European Research Council (ERC) under the European Union’s Horizon 2020 research and innovation programme (grant agreement No 863412). Computing resources were provided by TGCC and CINES under the allocation A0070407669 made by GENCI. A. P. acknowledges support by the National Science Foundation under Grant No. 2010145.

\end{acknowledgements}


\bibliographystyle{aa}
\bibliography{biblio}

\begin{thebibliography}{56}
\expandafter\ifx\csname natexlab\endcsname\relax\def\natexlab#1{#1}\fi

\bibitem[{{Acciari} {et~al.}(2009){Acciari}, {Aliu}, {Arlen}, {Bautista},
  {Beilicke}, {Benbow}, {Bradbury}, {Buckley}, {Bugaev}, {Butt},
  {et~al.}}]{Acciari_2009}
{Acciari}, V.~A., {Aliu}, E., {Arlen}, T., {et~al.} 2009, Science, 325, 444

\bibitem[{{Aharonian} {et~al.}(2009){Aharonian}, {Akhperjanian}, {Anton}, {de
  Almeida}, {Bazer-Bachi}, {Becherini}, {Behera}, {Benbow}, {Bernl{\"o}hr},
  {Boisson}, {Bochow}, {Borrel}, {Brion}, {Brucker}, {Brun}, {B{\"u}hler},
  {Bulik}, {B{\"u}sching}, {Boutelier}, {Chadwick}, {Charbonnier}, {Chaves},
  {Cheesebrough}, {Chounet}, {Clapson}, {Coignet}, {Dalton}, {Daniel},
  {Davids}, {Degrange}, {Deil}, {Dickinson}, {Djannati-Ata{\"\i}}, {Domainko},
  {Drury}, {Dubois}, {Dubus}, {Dyks}, {Dyrda}, {Egberts}, {Emmanoulopoulos},
  {Espigat}, {Farnier}, {Feinstein}, {Fiasson}, {F{\"o}rster}, {Fontaine},
  {F{\"u}{\ss}ling}, {Gabici}, {Gallant}, {G{\'e}rard}, {Giebels},
  {Glicenstein}, {Gl{\"u}ck}, {Goret}, {G{\"o}hring}, {Hauser}, {Hauser},
  {Heinz}, {Heinzelmann}, {Henri}, {Hermann}, {Hinton}, {Hoffmann}, {Hofmann},
  {Holleran}, {Hoppe}, {Horns}, {Jacholkowska}, {de Jager}, {Jahn}, {Jung},
  {Katarzy{\'n}ski}, {Katz}, {Kaufmann}, {Kendziorra}, {Kerschhaggl},
  {Khangulyan}, {Kh{\'e}lifi}, {Keogh}, {Klu{\'z}niak}, {Kneiske}, {Komin},
  {Kosack}, {Lamanna}, {Latham}, {Lenain}, {Lohse}, {Marand on}, {Martin},
  {Martineau-Huynh}, {Marcowith}, {Maurin}, {McComb}, {Medina}, {Moderski},
  {Moulin}, {Naumann-Godo}, {de Naurois}, {Nedbal}, {Nekrassov}, {Niemiec},
  {Nolan}, {Ohm}, {Olive}, {de O{\~n}a Wilhelmi}, {Orford}, {Ostrowski},
  {Panter}, {Arribas}, {Pedaletti}, {Pelletier}, {Petrucci}, {Pita},
  {P{\"u}hlhofer}, {Punch}, {Quirrenbach}, {Raubenheimer}, {Raue}, {Rayner},
  {Renaud}, {Rieger}, {Ripken}, {Rob}, {Rosier-Lees}, {Rowell}, {Rudak},
  {Rulten}, {Ruppel}, {Sahakian}, {Santangelo}, {Schlickeiser}, {Sch{\"o}ck},
  {Schr{\"o}der}, {Schwanke}, {Schwarzburg}, {Schwemmer}, {Shalchi}, {Sikora},
  {Skilton}, {Sol}, {Spangler}, {Stawarz}, {Steenkamp}, {Stegmann}, {Superina},
  {Szostek}, {Tam}, {Tavernet}, {Terrier}, {Tibolla}, {Tluczykont}, {van
  Eldik}, {Vasileiadis}, {Venter}, {Venter}, {Vialle}, {Vincent}, {Vink},
  {Vivier}, {V{\"o}lk}, {Volpe}, {Wagner}, {Ward}, {Zdziarski}, \&
  {Zech}}]{Aharonian_2009}
{Aharonian}, F., {Akhperjanian}, A.~G., {Anton}, G., {et~al.} 2009, \apjl, 695,
  L40

\bibitem[{{Aharonian} {et~al.}(2007){Aharonian}, {Akhperjanian}, {Bazer-Bachi},
  {Behera}, {Beilicke}, {Benbow}, {Berge}, {Bernl{\"o}hr}, {Boisson}, {Bolz},
  {Borrel}, {Boutelier}, {Braun}, {Brion}, {Brown}, {B{\"u}hler},
  {B{\"u}sching}, {Bulik}, {Carrigan}, {Chadwick}, {Clapson}, {Chounet},
  {Coignet}, {Cornils}, {Costamante}, {Degrange}, {Dickinson},
  {Djannati-Ata{\"\i}}, {Domainko}, {Drury}, {Dubus}, {Dyks}, {Egberts},
  {Emmanoulopoulos}, {Espigat}, {Farnier}, {Feinstein}, {Fiasson},
  {F{\"o}rster}, {Fontaine}, {Funk}, {Funk}, {F{\"u}{\ss}ling}, {Gallant},
  {Giebels}, {Glicenstein}, {Gl{\"u}ck}, {Goret}, {Hadjichristidis}, {Hauser},
  {Hauser}, {Heinzelmann}, {Henri}, {Hermann}, {Hinton}, {Hoffmann}, {Hofmann},
  {Holleran}, {Hoppe}, {Horns}, {Jacholkowska}, {de Jager}, {Kendziorra},
  {Kerschhaggl}, {Kh{\'e}lifi}, {Komin}, {Kosack}, {Lamanna}, {Latham}, {Le
  Gallou}, {Lemi{\`e}re}, {Lemoine-Goumard}, {Lenain}, {Lohse}, {Martin},
  {Martineau-Huynh}, {Marcowith}, {Masterson}, {Maurin}, {McComb}, {Moderski},
  {Moulin}, {de Naurois}, {Nedbal}, {Nolan}, {Olive}, {Orford}, {Osborne},
  {Ostrowski}, {Panter}, {Pedaletti}, {Pelletier}, {Petrucci}, {Pita},
  {P{\"u}hlhofer}, {Punch}, {Ranchon}, {Raubenheimer}, {Raue}, {Rayner},
  {Renaud}, {Ripken}, {Rob}, {Rolland}, {Rosier-Lees}, {Rowell}, {Rudak},
  {Ruppel}, {Sahakian}, {Santangelo}, {Saug{\'e}}, {Schlenker}, {Schlickeiser},
  {Schr{\"o}der}, {Schwanke}, {Schwarzburg}, {Schwemmer}, {Shalchi}, {Sol},
  {Spangler}, {Stawarz}, {Steenkamp}, {Stegmann}, {Superina}, {Tam},
  {Tavernet}, {Terrier}, {van Eldik}, {Vasileiadis}, {Venter}, {Vialle},
  {Vincent}, {Vivier}, {V{\"o}lk}, {Volpe}, {Wagner}, {Ward}, \&
  {Zdziarski}}]{Aharonian_2007}
{Aharonian}, F., {Akhperjanian}, A.~G., {Bazer-Bachi}, A.~R., {et~al.} 2007,
  \apjl, 664, L71

\bibitem[{{Aharonian} {et~al.}(2006){Aharonian}, {Akhperjanian}, {Bazer-Bachi},
  {Beilicke}, {Benbow}, {Berge}, {Bernl{\"o}hr}, {Boisson}, {Bolz}, {Borrel},
  {et~al.}}]{Aharonian_2006}
{Aharonian}, F., {Akhperjanian}, A.~G., {Bazer-Bachi}, A.~R., {et~al.} 2006,
  Science, 314, 1424

\bibitem[{{Albert} {et~al.}(2007){Albert}, {Aliu}, {Anderhub}, {Antoranz},
  {Armada}, {Baixeras}, {Barrio}, {Bartko}, {Bastieri}, {Becker}, {Bednarek},
  {Berger}, {Bigongiari}, {Biland}, {Bock}, {Bordas}, {Bosch-Ramon}, {Bretz},
  {Britvitch}, {Camara}, {Carmona}, {Chilingarian}, {Coarasa}, {Commichau},
  {Contreras}, {Cortina}, {Costado}, {Curtef}, {Danielyan}, {Dazzi}, {De
  Angelis}, {Delgado}, {de los Reyes}, {De Lotto}, {Domingo-Santamar{\'\i}a},
  {Dorner}, {Doro}, {Errando}, {Fagiolini}, {Ferenc}, {Fern{\'a}ndez}, {Firpo},
  {Flix}, {Fonseca}, {Font}, {Fuchs}, {Galante}, {Garc{\'\i}a-L{\'o}pez},
  {Garczarczyk}, {Gaug}, {Giller}, {Goebel}, {Hakobyan}, {Hayashida},
  {Hengstebeck}, {Herrero}, {H{\"o}hne}, {Hose}, {Hrupec}, {Hsu}, {Jacon},
  {Jogler}, {Kosyra}, {Kranich}, {Kritzer}, {Laille}, {Lindfors}, {Lombardi},
  {Longo}, {L{\'o}pez}, {L{\'o}pez}, {Lorenz}, {Majumdar}, {Maneva},
  {Mannheim}, {Mansutti}, {Mariotti}, {Mart{\'\i}nez}, {Mazin}, {Merck},
  {Meucci}, {Meyer}, {Miranda}, {Mirzoyan}, {Mizobuchi}, {Moralejo}, {Nieto},
  {Nilsson}, {Ninkovic}, {O{\~n}a-Wilhelmi}, {Otte}, {Oya}, {Paneque},
  {Panniello}, {Paoletti}, {Paredes}, {Pasanen}, {Pascoli}, {Pauss}, {Pegna},
  {Persic}, {Peruzzo}, {Piccioli}, {Prandini}, {Puchades}, {Raymers}, {Rhode},
  {Rib{\'o}}, {Rico}, {Rissi}, {Robert}, {R{\"u}gamer}, {Saggion}, {Saito},
  {S{\'a}nchez}, {Sartori}, {Scalzotto}, {Scapin}, {Schmitt}, {Schweizer},
  {Shayduk}, {Shinozaki}, {Shore}, {Sidro}, {Sillanp{\"a}{\"a}}, {Sobczynska},
  {Stamerra}, {Stark}, {Takalo}, {Tavecchio}, {Temnikov}, {Tescaro}, {Teshima},
  {Torres}, {Turini}, {Vankov}, {Vitale}, {Wagner}, {Wibig}, {Wittek},
  {Zandanel}, {Zanin}, \& {Zapatero}}]{Albert_2007}
{Albert}, J., {Aliu}, E., {Anderhub}, H., {et~al.} 2007, \apj, 669, 862

\bibitem[{{Aleksi{\'c}} {et~al.}(2014){Aleksi{\'c}}, {Ansoldi}, {Antonelli},
  {Antoranz}, {Babic}, {Bangale}, {Barrio}, {Gonz{\'a}lez}, {Bednarek},
  {Bernardini}, {Biasuzzi}, {Biland}, {Blanch}, {Bonnefoy}, {Bonnoli},
  {Borracci}, {Bretz}, {Carmona}, {Carosi}, {Colin}, {Colombo}, {Contreras},
  {Cortina}, {Covino}, {Da Vela}, {Dazzi}, {De Angelis}, {De Caneva}, {De
  Lotto}, {Wilhelmi}, {Mendez}, {Prester}, {Dorner}, {Doro}, {Einecke},
  {Eisenacher}, {Elsaesser}, {Fonseca}, {Font}, {Frantzen}, {Fruck}, {Galindo},
  {L{\'o}pez}, {Garczarczyk}, {Terrats}, {Gaug}, {Godinovi{\'c}}, {Mu{\~n}oz},
  {Gozzini}, {Hadasch}, {Hanabata}, {Hayashida}, {Herrera}, {Hildebrand },
  {Hose}, {Hrupec}, {Idec}, {Kadenius}, {Kellermann}, {Kodani}, {Konno},
  {Krause}, {Kubo}, {Kushida}, {La Barbera}, {Lelas}, {Lewandowska},
  {Lindfors}, {Lombardi}, {Longo}, {L{\'o}pez}, {L{\'o}pez-Coto},
  {L{\'o}pez-Oramas}, {Lorenz}, {Lozano}, {Makariev}, {Mallot}, {Maneva},
  {Mankuzhiyil}, {Mannheim}, {Maraschi}, {Marcote}, {Mariotti},
  {Mart{\'\i}nez}, {Mazin}, {Menzel}, {Mirand a}, {Mirzoyan}, {Moralejo},
  {Munar-Adrover}, {Nakajima}, {Niedzwiecki}, {Nilsson}, {Nishijima}, {Noda},
  {Orito}, {Overkemping}, {Paiano}, {Palatiello}, {Paneque}, {Paoletti},
  {Paredes}, {Paredes-Fortuny}, {Persic}, {Poutanen}, {Moroni}, {Prandini},
  {Puljak}, {Reinthal}, {Rhode}, {Rib{\'o}}, {Rico}, {Garcia}, {R{\"u}gamer},
  {Saito}, {Saito}, {Satalecka}, {Scalzotto}, {Scapin}, {Schultz}, {Schweizer},
  {Shore}, {Sillanp{\"a}{\"a}}, {Sitarek}, {Snidaric}, {Sobczynska}, {Spanier},
  {Stamatescu}, {Stamerra}, {Steinbring}, {Storz}, {Strzys}, {Takalo},
  {Takami}, {Tavecchio}, {Temnikov}, {Terzi{\'c}}, {Tescaro}, {Teshima},
  {Thaele}, {Tibolla}, {Torres}, {Toyama}, {Treves}, {Uellenbeck}, {Vogler},
  {Zanin}, {Kadler}, {Schulz}, {Ros}, {Bach}, {Krau{\ss}}, \&
  {Wilms}}]{Aleksic_2014}
{Aleksi{\'c}}, J., {Ansoldi}, S., {Antonelli}, L.~A., {et~al.} 2014, Science,
  346, 1080

\bibitem[{{Aliu} {et~al.}(2012){Aliu}, {Arlen}, {Aune}, {Beilicke}, {Benbow},
  {Bouvier}, {Bradbury}, {Buckley}, {Bugaev}, {Byrum}, {Cannon}, {Cesarini},
  {Ciupik}, {Collins-Hughes}, {Connolly}, {Cui}, {Dickherber}, {Duke},
  {Errando}, {Falcone}, {Finley}, {Finnegan}, {Fortson}, {Furniss}, {Galante},
  {Gall}, {Godambe}, {Griffin}, {Grube}, {Guenette}, {Gyuk}, {Hanna}, {Holder},
  {Huan}, {Hughes}, {Hui}, {Humensky}, {Imran}, {Kaaret}, {Karlsson},
  {Kertzman}, {Kieda}, {Krawczynski}, {Krennrich}, {Lang}, {LeBohec},
  {Madhavan}, {Maier}, {Majumdar}, {McArthur}, {McCann}, {Moriarty},
  {Mukherjee}, {Nu{\~n}ez}, {Ong}, {Orr}, {Otte}, {Park}, {Perkins}, {Pichel},
  {Pohl}, {Prokoph}, {Quinn}, {Ragan}, {Reyes}, {Reynolds}, {Roache}, {Rose},
  {Ruppel}, {Saxon}, {Schroedter}, {Sembroski}, {{\c{S}}ent{\"u}rk}, {Skole},
  {Staszak}, {Te{\v{s}}i{\'c}}, {Theiling}, {Thibadeau}, {Tsurusaki}, {Tyler},
  {Varlotta}, {Vassiliev}, {Vincent}, {Vivier}, {Wakely}, {Ward}, {Weekes},
  {Weinstein}, {Weisgarber}, {Williams}, \& {Zitzer}}]{Aliu_2012}
{Aliu}, E., {Arlen}, T., {Aune}, T., {et~al.} 2012, \apj, 746, 141

\bibitem[{{Bacchini} {et~al.}(2018){Bacchini}, {Ripperda}, {Chen}, \&
  {Sironi}}]{Bacchini_2018_1}
{Bacchini}, F., {Ripperda}, B., {Chen}, A.~Y., \& {Sironi}, L. 2018, apjs, 237,
  6

\bibitem[{{Bacchini} {et~al.}(2019){Bacchini}, {Ripperda}, {Porth}, \&
  {Sironi}}]{Bacchini_2018_2}
{Bacchini}, F., {Ripperda}, B., {Porth}, O., \& {Sironi}, L. 2019, apjs, 240,
  40

\bibitem[{{Bicak} \& {Janis}(1985)}]{Bicak_1985}
{Bicak}, J. \& {Janis}, V. 1985, \mnras, 212, 899

\bibitem[{{Blandford} \& {Znajek}(1977)}]{Blandford_1977}
{Blandford}, R.~D. \& {Znajek}, R.~L. 1977, mnras, 179, 433

\bibitem[{{Carter}(1968)}]{Carter_1968}
{Carter}, B. 1968, Physical Review, 174, 1559

\bibitem[{{Cerutti} {et~al.}(2015){Cerutti}, {Philippov}, {Parfrey}, \&
  {Spitkovsky}}]{Cerutti_2015}
{Cerutti}, B., {Philippov}, A., {Parfrey}, K., \& {Spitkovsky}, A. 2015, mnras,
  448, 606

\bibitem[{{Cerutti} {et~al.}(2016){Cerutti}, {Philippov}, \&
  {Spitkovsky}}]{Cerutti_2016}
{Cerutti}, B., {Philippov}, A.~A., \& {Spitkovsky}, A. 2016, \mnras, 457, 2401

\bibitem[{Cerutti {et~al.}(2012)Cerutti, Uzdensky, \& Begelman}]{Cerutti_2012}
Cerutti, B., Uzdensky, D.~A., \& Begelman, M.~C. 2012, The Astrophysical
  Journal, 746, 148

\bibitem[{{Cerutti} {et~al.}(2013){Cerutti}, {Werner}, {Uzdensky}, \&
  {Begelman}}]{Cerutti_2013}
{Cerutti}, B., {Werner}, G.~R., {Uzdensky}, D.~A., \& {Begelman}, M.~C. 2013,
  \apj, 770, 147

\bibitem[{{Chen} \& {Yuan}(2020)}]{Chen_2020}
{Chen}, A.~Y. \& {Yuan}, Y. 2020, \apj, 895, 121

\bibitem[{{Christie} {et~al.}(2019){Christie}, {Petropoulou}, {Sironi}, \&
  {Giannios}}]{Christie_2019}
{Christie}, I.~M., {Petropoulou}, M., {Sironi}, L., \& {Giannios}, D. 2019,
  \mnras, 482, 65

\bibitem[{{Crinquand} {et~al.}(2020){Crinquand}, {Cerutti}, {Philippov},
  {Parfrey}, \& {Dubus}}]{Crinquand_2020}
{Crinquand}, B., {Cerutti}, B., {Philippov}, A., {Parfrey}, K., \& {Dubus}, G.
  2020, \prl, 124, 145101

\bibitem[{{Cunningham} \& {Bardeen}(1973)}]{Cunningham_1973}
{Cunningham}, C.~T. \& {Bardeen}, J.~M. 1973, \apj, 183, 237

\bibitem[{{Dexter} \& {Agol}(2009)}]{Dexter_2009}
{Dexter}, J. \& {Agol}, E. 2009, \apj, 696, 1616

\bibitem[{{Dodin} \& {Fisch}(2010)}]{Dodin_2010}
{Dodin}, I.~Y. \& {Fisch}, N.~J. 2010, Physics of Plasmas, 17, 112118

\bibitem[{{Event Horizon Telescope Collaboration} {et~al.}(2019){Event Horizon
  Telescope Collaboration}, {Akiyama}, {Alberdi}, {Alef}, {Asada}, {Azulay},
  {Baczko}, {Ball}, {Balokovi{\'c}}, {Barrett}, {et~al.}}]{EHT_1}
{Event Horizon Telescope Collaboration}, {Akiyama}, K., {Alberdi}, A., {et~al.}
  2019, \apj, 875, L1

\bibitem[{{Font} {et~al.}(1999){Font}, {Ib{\'a}{\~n}ez}, \&
  {Papadopoulos}}]{Font_1999}
{Font}, J.~A., {Ib{\'a}{\~n}ez}, J.~M., \& {Papadopoulos}, P. 1999, \mnras,
  305, 920

\bibitem[{{Gould} \& {Schr{\'e}der}(1967)}]{Gould_1967}
{Gould}, R.~J. \& {Schr{\'e}der}, G.~P. 1967, Physical Review, 155, 1404

\bibitem[{{Gralla} {et~al.}(2018){Gralla}, {Lupsasca}, \&
  {Strominger}}]{Gralla_2018}
{Gralla}, S.~E., {Lupsasca}, A., \& {Strominger}, A. 2018, \mnras, 475, 3829

\bibitem[{{Gravity Collaboration} {et~al.}(2018){Gravity Collaboration},
  {Abuter}, {Amorim}, {Baub{\"o}ck}, {Berger}, {Bonnet}, {Brand ner},
  {Cl{\'e}net}, {Coud{\'e} Du Foresto}, {de Zeeuw}, {Deen}, {Dexter}, {Duvert},
  {Eckart}, {Eisenhauer}, {F{\"o}rster Schreiber}, {Garcia}, {Gao}, {Gendron},
  {Genzel}, {Gillessen}, {Guajardo}, {Habibi}, {Haubois}, {Henning}, {Hippler},
  {Horrobin}, {Huber}, {Jim{\'e}nez-Rosales}, {Jocou}, {Kervella}, {Lacour},
  {Lapeyr{\`e}re}, {Lazareff}, {Le Bouquin}, {L{\'e}na}, {Lippa}, {Ott},
  {Panduro}, {Paumard}, {Perraut}, {Perrin}, {Pfuhl}, {Plewa}, {Rabien},
  {Rodr{\'\i}guez-Coira}, {Rousset}, {Sternberg}, {Straub}, {Straubmeier},
  {Sturm}, {Tacconi}, {Vincent}, {von Fellenberg}, {Waisberg}, {Widmann},
  {Wieprecht}, {Wiezorrek}, {Woillez}, \& {Yazici}}]{Gravity_2018}
{Gravity Collaboration}, {Abuter}, R., {Amorim}, A., {et~al.} 2018, \aap, 618,
  L10

\bibitem[{{H.~E.~S.~S. Collaboration} {et~al.}(2017){H.~E.~S.~S.
  Collaboration}, {Abdalla}, {Abramowski}, {Aharonian}, {Ait Benkhali},
  {Akhperjanian}, {Andersson}, {Ang{\"u}ner}, {Arrieta}, {Aubert}, {Backes},
  {Balzer}, {Barnard}, {Becherini}, {Becker Tjus}, {Berge}, {Bernhard},
  {Bernl{\"o}hr}, {Blackwell}, {B{\"o}ttcher}, {Boisson}, {Bolmont}, {Bordas},
  {Bregeon}, {Brun}, {Brun}, {Bryan}, {Bulik}, {Capasso}, {Carr}, {Casanova},
  {Cerruti}, {Chakraborty}, {Chalme-Calvet}, {Chaves}, {Chen}, {Chevalier},
  {Chr{\'e}tien}, {Colafrancesco}, {Cologna}, {Condon}, {Conrad}, {Cui},
  {Davids}, {Decock}, {Degrange}, {Deil}, {Devin}, {deWilt}, {Dirson},
  {Djannati-Ata{\"\i}}, {Domainko}, {Donath}, {Drury}, {Dubus}, {Dutson},
  {Dyks}, {Edwards}, {Egberts}, {Eger}, {Ernenwein}, {Eschbach}, {Farnier},
  {Fegan}, {Fernand es}, {Fiasson}, {Fontaine}, {F{\"o}rster}, {Funk},
  {F{\"u}{\ss}ling}, {Gabici}, {Gajdus}, {Gallant}, {Garrigoux}, {Giavitto},
  {Giebels}, {Glicenstein}, {Gottschall}, {Goyal}, {Grondin}, {Hadasch},
  {Hahn}, {Haupt}, {Hawkes}, {Heinzelmann}, {Henri}, {Hermann}, {Hervet},
  {Hinton}, {Hofmann}, {Hoischen}, {Holler}, {Horns}, {Ivascenko},
  {Jacholkowska}, {Jamrozy}, {Janiak}, {Jankowsky}, {Jankowsky}, {Jingo},
  {Jogler}, {Jouvin}, {Jung-Richardt}, {Kastendieck}, {Katarzy{\'n}ski},
  {Katz}, {Kerszberg}, {Kh{\'e}lifi}, {Kieffer}, {King}, {Klepser}, {Klochkov},
  {Klu{\'z}niak}, {Kolitzus}, {Komin}, {Kosack}, {Krakau}, {Kraus}, {Krayzel},
  {Kr{\"u}ger}, {Laffon}, {Lamanna}, {Lau}, {Lees}, {Lefaucheur}, {Lefranc},
  {Lemi{\`e}re}, {Lemoine-Goumard}, {Lenain}, {Leser}, {Lohse}, {Lorentz},
  {Liu}, {L{\'o}pez-Coto}, {Lypova}, {Marandon}, {Marcowith}, {Mariaud},
  {Marx}, {Maurin}, {Maxted}, {Mayer}, {Meintjes}, {Meyer}, {Mitchell},
  {Moderski}, {Mohamed}, {Mohrmann}, {Mor{\r{a}}}, {Moulin}, {Murach}, {de
  Naurois}, {Niederwanger}, {Niemiec}, {Oakes}, {O'Brien}, {Odaka}, {{\"O}ttl},
  {Ohm}, {Ostrowski}, {Oya}, {Padovani}, {Panter}, {Parsons}, {Pekeur},
  {Pelletier}, {Perennes}, {Petrucci}, {Peyaud}, {Piel}, {Pita}, {Poon},
  {Prokhorov}, {Prokoph}, {P{\"u}hlhofer}, {Punch}, {Quirrenbach}, {Raab},
  {Reimer}, {Reimer}, {Renaud}, {de los Reyes}, {Rieger}, {Romoli},
  {Rosier-Lees}, {Rowell}, {Rudak}, {Rulten}, {Sahakian}, {Salek}, {Sanchez},
  {Santangelo}, {Sasaki}, {Schlickeiser}, {Sch{\"u}ssler}, {Schulz},
  {Schwanke}, {Schwemmer}, {Settimo}, {Seyffert}, {Shafi}, {Shilon}, {Simoni},
  {Sol}, {Spanier}, {Spengler}, {Spies}, {Stawarz}, {Steenkamp}, {Stegmann},
  {Stinzing}, {Stycz}, {Sushch}, {Tavernet}, {Tavernier}, {Taylor}, {Terrier},
  {Tibaldo}, {Tiziani}, {Tluczykont}, {Trichard}, {Tuffs}, {Uchiyama}, {van der
  Walt}, {van Eldik}, {van Rensburg}, {van Soelen}, {Vasileiadis}, {Veh},
  {Venter}, {Viana}, {Vincent}, {Vink}, {Voisin}, {V{\"o}lk}, {Vuillaume},
  {Wadiasingh}, {Wagner}, {Wagner}, {Wagner}, {White}, {Wierzcholska},
  {Willmann}, {W{\"o}rnlein}, {Wouters}, {Yang}, {Zabalza}, {Zaborov},
  {Zacharias}, {Zdziarski}, {Zech}, {Zefi}, {Ziegler}, \&
  {{\.Z}ywucka}}]{Hess_2017}
{H.~E.~S.~S. Collaboration}, {Abdalla}, H., {Abramowski}, A., {et~al.} 2017,
  \aap, 598, A39

\bibitem[{{Hirotani} \& {Pu}(2016)}]{Hirotani2016}
{Hirotani}, K. \& {Pu}, H.-Y. 2016, \apj, 818, 50

\bibitem[{Hughes {et~al.}(1994)Hughes, Keeton, Walker, Walsh, Shapiro, \&
  Teukolsky}]{Hughes_1994}
Hughes, S.~A., Keeton, C.~R., Walker, P., {et~al.} 1994, Phys. Rev. D, 49, 4004

\bibitem[{{Huppenkothen} {et~al.}(2019){Huppenkothen}, {Bachetti}, {Stevens},
  {Migliari}, {Balm}, {Hammad}, {Khan}, {Mishra}, {Rashid}, {Sharma}, {Martinez
  Ribeiro}, \& {Valles Blanco}}]{Huppenkothen_2019}
{Huppenkothen}, D., {Bachetti}, M., {Stevens}, A.~L., {et~al.} 2019, \apj, 881,
  39

\bibitem[{{Jacquemin-Ide} {et~al.}(2020){Jacquemin-Ide}, {Lesur}, \&
  {Ferreira}}]{Jacquemin_2020}
{Jacquemin-Ide}, J., {Lesur}, G., \& {Ferreira}, J. 2020, arXiv e-prints,
  arXiv:2011.14782

\bibitem[{{Kagan} {et~al.}(2015){Kagan}, {Sironi}, {Cerutti}, \&
  {Giannios}}]{Kagan_2015}
{Kagan}, D., {Sironi}, L., {Cerutti}, B., \& {Giannios}, D. 2015, \ssr, 191,
  545

\bibitem[{{Katsoulakos} \& {Rieger}(2018)}]{Katsoulakos_2018}
{Katsoulakos}, G. \& {Rieger}, F.~M. 2018, \apj, 852, 112

\bibitem[{{Kisaka} {et~al.}(2020){Kisaka}, {Levinson}, \& {Toma}}]{Kisaka_2020}
{Kisaka}, S., {Levinson}, A., \& {Toma}, K. 2020, arXiv e-prints,
  arXiv:2007.02838

\bibitem[{{Komissarov}(2004)}]{Komissarov_2004a}
{Komissarov}, S.~S. 2004, \mnras, 350, 427

\bibitem[{{Komissarov} \& {McKinney}(2007)}]{Komissarov_2007}
{Komissarov}, S.~S. \& {McKinney}, J.~C. 2007, mnras, 377, L49

\bibitem[{{Levinson}(2000)}]{Levinson_2000}
{Levinson}, A. 2000, \prl, 85, 912

\bibitem[{{Levinson} \& {Cerutti}(2018)}]{Levinson_2018}
{Levinson}, A. \& {Cerutti}, B. 2018, aap, 616, A184

\bibitem[{{Levinson} \& {Rieger}(2011)}]{Levinson_2011}
{Levinson}, A. \& {Rieger}, F. 2011, \apj, 730, 123

\bibitem[{{Lubow} {et~al.}(1994){Lubow}, {Papaloizou}, \&
  {Pringle}}]{Lubow_1994}
{Lubow}, S.~H., {Papaloizou}, J.~C.~B., \& {Pringle}, J.~E. 1994, \mnras, 267,
  235

\bibitem[{{MacDonald} \& {Thorne}(1982)}]{MacDonald_1982}
{MacDonald}, D. \& {Thorne}, K.~S. 1982, mnras, 198, 345

\bibitem[{Mahlmann {et~al.}(2020)Mahlmann, Levinson, \& Aloy}]{Mahlmann_2020}
Mahlmann, J.~F., Levinson, A., \& Aloy, M.~A. 2020, Monthly Notices of the
  Royal Astronomical Society, 494, 4203–4225

\bibitem[{{McKinney} {et~al.}(2012){McKinney}, {Tchekhovskoy}, \&
  {Blandford}}]{McKinney_2012}
{McKinney}, J.~C., {Tchekhovskoy}, A., \& {Blandford}, R.~D. 2012, \mnras, 423,
  3083

\bibitem[{{Mehlhaff} {et~al.}(2020){Mehlhaff}, {Werner}, {Uzdensky}, \&
  {Begelman}}]{Mehlhaff_2020}
{Mehlhaff}, J.~M., {Werner}, G.~R., {Uzdensky}, D.~A., \& {Begelman}, M.~C.
  2020, \mnras, 498, 799

\bibitem[{{Narayan} {et~al.}(2003){Narayan}, {Igumenshchev}, \&
  {Abramowicz}}]{Narayan_2003}
{Narayan}, R., {Igumenshchev}, I.~V., \& {Abramowicz}, M.~A. 2003, \pasj, 55,
  L69

\bibitem[{Parfrey {et~al.}(2015)Parfrey, Giannios, \&
  Beloborodov}]{Parfrey_2015}
Parfrey, K., Giannios, D., \& Beloborodov, A.~M. 2015, Monthly Notices of the
  Royal Astronomical Society, 446, L61–L65

\bibitem[{{Parfrey} {et~al.}(2019){Parfrey}, {Philippov}, \&
  {Cerutti}}]{Parfrey_2019}
{Parfrey}, K., {Philippov}, A., \& {Cerutti}, B. 2019, Physical Review Letters,
  122, 035101

\bibitem[{{Rieger} \& {Aharonian}(2012)}]{Rieger_2012}
{Rieger}, F.~M. \& {Aharonian}, F. 2012, Modern Physics Letters A, 27, 1230030

\bibitem[{{Ripperda} {et~al.}(2020){Ripperda}, {Bacchini}, \&
  {Philippov}}]{Ripperda_2020}
{Ripperda}, B., {Bacchini}, F., \& {Philippov}, A.~A. 2020, \apj, 900, 100

\bibitem[{{Spitkovsky}(2006)}]{Spitkovsky_2006}
{Spitkovsky}, A. 2006, \apjl, 648, L51

\bibitem[{{Tchekhovskoy} {et~al.}(2010){Tchekhovskoy}, {Narayan}, \&
  {McKinney}}]{Tchekhovskoy_2010}
{Tchekhovskoy}, A., {Narayan}, R., \& {McKinney}, J.~C. 2010, \apj, 711, 50

\bibitem[{Uzdensky {et~al.}(2010)Uzdensky, Loureiro, \&
  Schekochihin}]{Uzdensky_2010}
Uzdensky, D.~A., Loureiro, N.~F., \& Schekochihin, A.~A. 2010, 105, 235002

\bibitem[{{Wald}(1974)}]{Wald_1974}
{Wald}, R. 1974, \prd, 10, 1680

\bibitem[{{Werner} {et~al.}(2018){Werner}, {Uzdensky}, {Begelman}, {Cerutti},
  \& {Nalewajko}}]{Werner_2018}
{Werner}, G.~R., {Uzdensky}, D.~A., {Begelman}, M.~C., {Cerutti}, B., \&
  {Nalewajko}, K. 2018, \mnras, 473, 4840

\bibitem[{{Yee}(1966)}]{Yee_1966}
{Yee}, K. 1966, IEEE Transactions on Antennas and Propagation, 14, 302

\end{thebibliography}

\begin{appendix}

\section{Light curve reconstruction} \label{appendix}

Let us denote by $p^\mu$ the $4$-momentum of a photon with initial coordinates $(t_0, r_0, \theta_0, \varphi_0)$. The motion of the photon is specified by several constants of motion: the energy at infinity $\mathcal{E}=-p_t$, the component of the angular momentum relative to the spin axis of the black hole $L=-p_\varphi$, and the Carter constant $Q=p_\theta^2 - \cos^2{\theta} \left( a^2 \mathcal{E}^2 - L^2 / \sin^2{\theta} \right)$~\citep{Carter_1968}. We also define the re-scaled quantities $l=L/\mathcal{E}$ and $q^2=Q/\mathcal{E}^2$. The photon trajectory is entirely determined by its initial position, the parameters $l$ and $q^2$, and two choices of sign in its initial radial and polar directions. It is independent of the photon energy. The PIC code \texttt{Zeltron}, at every time step, saves the quantities $q$, $l$, $r_0$, $\theta_0$, $t_0$, $\varphi_0$, and $E$, as well as the initial $\mathrm{sgn}(\dot{r})$ and $\mathrm{sgn}(\dot{\theta})$. The code \texttt{geokerr} is then used to integrate every saved geodesics up to $r=r_\mathrm{out}$, and outputs the final photon coordinates $(t_\mathrm{f},r_\mathrm{f},\theta_\mathrm{f},\varphi_\mathrm{f})$. If the photon ends up at infinity, the spherical components of the normalized $3$-velocity can be reconstructed~\citep{Cunningham_1973,Gralla_2018}:

\begin{align}
    v^\theta & =\pm \dfrac{1}{r_\mathrm{f}} \sqrt{q^2 + a^2 \cos^2{\theta_\mathrm{f}} - l^2 \cos^2{\theta_\mathrm{f}} / \sin^2{\theta_\mathrm{f}}}; \\
    v^\varphi & = \dfrac{l}{r_\mathrm{f} \sin{\theta_\mathrm{f}}} ; \\
    v^r & = \sqrt{1 - \left( v^\theta \right)^2 - \left( v^\varphi \right)^2}.
\end{align}

The photons are then collected at infinity. Let us define a Cartesian basis $(\vec{e_x}, \vec{e_y}, \vec{e_z})$, where $\vec{e_z}$ is aligned with the spin axis of the black hole and where $\varphi$ is measured in the $(\vec{e_x},\vec{e_y})$ plane with respect to $\vec{e_x}$. On this basis, the final direction of a photon is defined by its co-latitude $\alpha$ and its azimuth $\psi$. Our simulations being axisymmetric, we assume that observers are solely characterized by a viewing angle $\alpha_\mathrm{obs}$, and are able to collect all photons whose final direction is $\alpha = \alpha_\mathrm{obs}$, irrespective of $\psi$. The angle $\alpha$ is determined by

\begin{equation}
    \cos{\alpha} = \vec{v} \cdot \vec{e_z} = \cos{\theta_\mathrm{f}} \, v^r - \sin{\theta_\mathrm{f}} \, v^\theta,
\end{equation}
whereas $\psi$ is determined $\cos{\psi} = v^x / \sqrt{\left(v^x \right)^2 + \left( v^y \right)^2}$ and $\sin{\psi} = v^y / \sqrt{\left(v^x \right)^2 + \left( v^y \right)^2}$ (with $v^x=\vec{v}\cdot \vec{e_x}$ and $v^y=\vec{v}\cdot \vec{e_y}$). The eventual time delay for a photon at position $(r_\mathrm{r},\theta_\mathrm{f}, \varphi_\mathrm{f})$, propagating toward an observer at $(\alpha_\mathrm{obs},\psi)$, is computed as~\citep{Cerutti_2016}
\begin{equation}
    c \, \delta t = r_\mathrm{out} - r_\mathrm{f} \left( \sin{\alpha_\mathrm{obs}} \sin{\theta_\mathrm{f}} \cos{(\varphi_\mathrm{f} - \psi)} + \cos{\alpha_\mathrm{obs}} \cos{\theta_\mathrm{f}} \right).
\end{equation}
The total time delay, from emission to collection by an observer, is $t_\mathrm{delay} = t_\mathrm{f}- t_0 + \delta t$. All in all, a photon arriving at $r_\mathrm{out}$ with an angle $\theta_\mathrm{f}$ at a time $t_\mathrm{f}$ contributes to the luminous intensity per unit solid angle, at the time $t_0 + t_\mathrm{delay}$ and in the direction $\theta_\mathrm{f}$:
\begin{equation}
    \ddroit{L_\gamma}{\Omega}  = \dfrac{\diff{\mathcal{E}}}{2 \pi \sin \theta \diff{\theta} \diff{t}} .
\end{equation}
We do not keep any spectral information because of the low separation of scales: the fluxes we compute are integrated over all photon energies.

\texttt{Zeltron} uses spherical Kerr-Schild coordinates $(t,r,\theta,\varphi)$, whereas \texttt{geokerr} uses spherical Boyer-Lindquist coordinates $(\tilde{t},\tilde{r},\tilde{\theta},\tilde{\varphi})$. The coordinate transformation from Boyer-Lindquist to Kerr-Schild reads~\citep{Komissarov_2004a,Font_1999}
\begin{align}
    \mathrm{d}t & = \mathrm{d}\tilde{t} + \dfrac{2 r}{r^2 + a^2 - 2 r} \mathrm{d}r, \\
    \mathrm{d}\varphi & = \mathrm{d}\tilde{\varphi} + \dfrac{a}{r^2 + a^2 - 2 r} \mathrm{d}r,
\end{align}
along with $r=\tilde{r}$ and $\theta=\tilde{\theta}$. This yields, in integrated form,
\begin{align}
    t & = \tilde{t} + \ln{\left(r^2-2r+a^2 \right)} + \dfrac{1}{\sqrt{1-a^2}} \ln \left( \dfrac{r- \left( 1+\sqrt{1-a^2} \right)}{r- \left( 1-\sqrt{1-a^2} \right)} \right), \\
    \varphi & = \tilde{\varphi} + \dfrac{a}{2 \sqrt{1-a^2}} \ln \left( \dfrac{r-\left( 1+\sqrt{1-a^2} \right)}{r-\left( 1-\sqrt{1-a^2} \right)} \right).
\end{align}
The $t$ and $\varphi$ coordinates of emission are transformed from Kerr-Schild to Boyer-Lindquist before being dumped by \texttt{Zeltron}. The covariant components $x_t$, $x_\theta$, and $x_\varphi$ of any one-form $x_\mu$ are left unchanged by this transformation, so all constants of the motion have the same expression  in both coordinate systems.

From the values of $q$, $l$, and $r_0$ for a given photon, it is possible to determine whether this photon will end up in the black hole or at infinity, by solving for the roots of the radial potential $R(r)=(r^2 + a^2 - a l)^2 - (r^2 +a^2 - 2 r) (q^2 + (a-l)^2)$ \citep{Gralla_2018}. We perform this test so that only the photons that make it to infinity and actually contribute to the light curve are saved. For computational reasons, we only keep a fixed fraction of these photons (between $5\%$ and $10\%$), which is sufficient to reconstruct the angularly resolved light curve. 

We ran several sanity checks after coupling \texttt{geokerr} and \texttt{Zeltron}. While \texttt{geokerr} conserves the constants of the motion by construction, \texttt{Zeltron} does not. First we checked that for a large sample of photons whose trajectories are integrated by \texttt{Zeltron}, $\mathcal{E}$ and $Q$ are conserved to a relative accuracy of $10^{-10}$. Knowing that \texttt{Zeltron's} integrator is reliable, we compared trajectories integrated by the two codes, and checked that they matched to a similar accuracy, regardless of the complexity of the geodesic (number of turning points in $r$ or $\theta$) or the number of points computed on the geodesic by \texttt{geokerr}. 

\end{appendix}

\end{document}